%% file: BPP_Journal_Arxive.tex
\documentclass[journal,draftclsnofoot,onecolumn,12pt]{IEEEtran}

\usepackage{amsthm,amssymb,graphicx,multirow,amsmath,color,amsfonts}
\usepackage[update,prepend]{epstopdf}
\usepackage[noadjust]{cite}
\usepackage[latin1]{inputenc}
\usepackage{tikz}
\usetikzlibrary{arrows,calc}		
\usepackage{bbm} 
\usepackage{pdfpages}
\usepackage{tabulary}
\usepackage{multirow}
\usepackage{comment}

\def\chb#1{{\color{black} #1}}

\include{notation}

\allowdisplaybreaks 

\begin{document}
\graphicspath{{./Figures/}}
\title{
 Fundamentals of Modeling Finite Wireless Networks using Binomial Point Process}
\author{Mehrnaz Afshang and Harpreet S. Dhillon
\thanks{The authors are with Wireless@VT, Department of ECE, Virgina Tech, Blacksburg, VA, USA. Email: \{mehrnaz, hdhillon\}@vt.edu. The support of the US NSF (Grant CCF-1464293) is gratefully acknowledged.} 
\thanks{\chb{This paper will be presented in part at the IEEE SPAWC,  Edinburgh, UK, 2016 \cite{afshang2016optimal1}. \hfill Last updated: \today.}}}

\maketitle
\vspace{-.6cm}

\begin{abstract}

Modeling the locations of nodes as a uniform binomial point process (BPP), we present a generic mathematical framework to characterize the performance of an  arbitrarily-located reference receiver in a finite wireless network. Different from most of the prior works on the analysis of BPP where the serving \chb{transmitter (TX) node} is located at the fixed distance from the reference receiver, we consider two general \chb{TX-selection} policies: i)  \chb{\em uniform TX-selection}: the serving node is chosen uniformly at random amongst transmitting nodes, and ii) \chb{\em $k$-closest TX-selection}: the serving node is the $k^{th}$ closest node out of transmitting nodes to the reference receiver, where  $k=1$ models the so called {\em nearest neighbor} connectivity.
The key intermediate step in our analysis is the derivation of a new set of distance distributions that lead not only to the tractable analysis of coverage probability but also enable the analyses of  wide range of classical and currently trending problems 
 in wireless networks, where transmitting nodes are confined in a finite area.  
  In particular,  using the new set of distance distributions, we first study the classical problem of diversity loss due to  signal-to-interference ($\sir$) correlation under selection combining scheme. Our analysis reveals that ignoring  the $\sir$ correlation caused
by the common locations of interfering nodes significantly overestimates the performance of selection combining scheme.
   Second,  we characterize \chb{network spectral efficiency} (${\tt NSE}$) of a given finite network for the
two \chb{TX-selection} policies. Our analysis shows that the optimal number of simultaneously active nodes that maximizes ${\tt NSE}$ strongly depends on \chb{TX-selection} policy. Third,  using the new coverage probability result, we evaluate the optimal caching probability of the popular content to maximize the
total hit probability.  Our analysis demonstrates that
optimal caching probability is a strong  function of the number of
simultaneously active nodes in the network.

\end{abstract}
\begin{IEEEkeywords}
Binomial point process, finite wireless network, $k$-coverage analysis, optimal cache placement, selection combining scheme, stochastic geometry.
\end{IEEEkeywords}

\section{Introduction} \label{sec:intro}

{{\IEEEPARstart {P}{roper} spatial modeling of wireless networks is important for their accurate visualization, design, and performance analysis. Irrespective of the wireless network topology, homogeneous (often infinite) Poisson Point Process (PPP) is the most popular choice due to its simplicity and tractability \cite{Andrewsprimer2010,elsawy2013stochastic,andrews2016primer}. Despite its relevance in modeling large-scale random networks, it cannot be used to model finite network with a given number of nodes. This scenario is becoming mainstream with the popularity of millimeter wave (mmWave) communications. A popular choice in such cases is a BPP~\cite{haenggi2012stochastic}. }
 
 The performance analysis of  a wireless network modeled by a BPP is significantly more challenging \chb{compared to an infinite PPP}  due to three main reasons.  \chb{First, the performance is location dependent.} For \chb{instance}, the  aggregate interference and  the $\sir$  seen at the center of network are different from that of network boundary. \chb{Second},  the distances from all transmitting nodes to \chb{an}  arbitrarily located reference receiver are correlated due to \chb{the} common distance from the reference receiver  to the \chb{{\em center} of the network}. \chb{As discussed next, these challenges have already been addressed in the literature under the assumption that the receiving nodes are located at a given fixed distance from their corresponding serving nodes}~\cite{srinivasa2010distance,torrieri2012outage,venugopal2015interference,venugopal2015analysis, guo2014outage,guo2015performance,valenti2014direct}. The third challenge is induced by the selection of the serving node from  the set of finite nodes confined in the finite region. For example, if the reference receiver is served by \chb{its} nearest  transmitting \chb{node}, the  distribution of \chb{the} point process after removing serving node is not the same as that of the original BPP. The selection of serving node from the point  process is  important  for  \chb{the} modeling and analysis of  cellular networks and  several emerging applications  of wireless networks.
 \chb{This challenge has not yet been addressed comprehensively in the literature and is the main focus of this paper. In particular, we develop new tools to facilitate performance analysis of finite wireless networks under generic TX selection policies. We also present several applications of the proposed analytic tools to both classic and emerging problems in wireless networks.}
\subsection{Motivation and Related Work}
Existing works on modeling and analysis of   \chb{\em finite} wireless \chb{networks} have taken two main directions.  The first considers
a relatively simple setup  where the reference receiver is located at   the center of circular or annular region \cite{srinivasa2010distance,torrieri2012outage,venugopal2015interference,venugopal2015analysis,DhiVishnu2016UAV}. {This simple setup is  widely used in the   analysis \chb{of}  metrics defined in terms of  $\sir$ distribution such as  outage probability and transmission capacity of wireless {\em ad hoc} \cite{srinivasa2010distance,torrieri2012outage} and \chb{mmWave}  communication networks \cite{venugopal2015interference,venugopal2015analysis}. } \chb{Second, which can actually be treated as an extension of the first, is to consider more general setups with an arbitrarily located reference receiver in arbitrarily-shaped finite wireless networkss~\cite{guo2014outage,guo2015performance,valenti2014direct}}. 
  However, all these  works~\cite{srinivasa2010distance,torrieri2012outage,venugopal2015interference,venugopal2015analysis, guo2014outage,guo2015performance,valenti2014direct} study the performance of  a given link where the transmitter is located at a fixed distance from the receiver. \chb{ While} the  fixed link distance analysis  provides \chb{some useful insights} on  the performance of  finite networks, \chb{it is not always accurate}. For example, users are  typically associated to the base station (BS) that provides maximum average  received power  in the existing cellular network~\cite{andrews2011tractable}. This maximum average  received power association  can be interpreted as {nearest neighbor connectivity} wherein the user connects to the  closest BS in a single tier network. As a direct consequence,  interfering BSs must be farther than serving BS to the reference receiver. \chb{While} this effect can be easily captured \chb{when} the nodes are distributed according to an infinite PPP \cite{andrews2011tractable}, the  characterization of the  {nearest neighbor connectivity} in finite networks is challenging. \chb{It is studied} in~\cite{BananiFinite} by approximating the 
 $\sir$ conditioned on the location of serving node as  a lognormal random variable.  However, the  exact characterization of interference field in finite networks (e.g., hotspots and indoor network) where the serving node is  a part of the point process is still an open problem. This problem gets more challenging where the serving node is  the $k^{th}$ closest node to the reference receiver ($k=1$ is the closest).  The exact characterization of the  performance of an arbitrarily-located reference receiver    under two generic \chb{TX-selection}  policies where the serving node is a part of the transmitting node process is the main focus of this paper. More details are provided next.

\subsection{Contributions and Outcomes}
 {\em Modeling and analysis of finite wireless network.}
 We develop a comprehensive framework  for the performance evaluation of  {\em finite} wireless networks. In particular, we model the locations of   nodes as a uniform BPP to study the performance of an arbitrarily-located reference receiver  
under two \chb{TX-selection}  policies: i)  uniform TX-selection policy where  the serving node is  chosen uniformly at random
 amongst  set of transmitting nodes, and ii) \chb{$k$-closest TX-selection} policy where   the serving node is the $k^{th}$ closest node out of transmitting nodes to the reference receiver.  It is worth noting that \chb{uniform TX-selection} policy is more relevant to {\em ad hoc} setup where the reference receiver may connect to  one of the nodes at random.  \chb{On the other hand,  \chb{$k$-closest TX-selection} policy is more relevant for the performance analysis of cellular networks, especially its applications to localization~\cite{schloemann2015towards} and geographic caching~\cite{BlaszczyszynG14}.} The coverage probability of \chb{$k$-closest TX-selection} policy   is  analogous to    {$k$-coverage result} of~\cite{H.P.Keelerkcoverage}  for infinite PPP. {As discussed next}, the performance analysis of these two setups where the serving node is  a part of the point process bring forth new technical challenges, e.g., the need to characterize
the distribution of distances from reference receiver to the interfering and serving nodes.
 
{\em Coverage probability analysis. }
 We first derive the ``exact'' expression for coverage probability of an arbitrarily-located reference receiver under two \chb{TX-selection} policies explained above. It is
then specialized and extended to two cases of interest: i) central receiver: the receiver is located at center of circular region, and ii) {random receiver: the receiver is a randomly chosen receiver out of receiving nodes.} To perform this analysis, we characterize  distance distributions from an arbitrarily-located reference receiver to the serving and interfering nodes as a key intermediate step in the two TX-selection policies. 
The  exact analysis of  \chb{$k$-closest TX-selection} policy  requires more careful analysis of the interference field.
In particular,  we prove that the distances from interfering nodes conditioned on the location of serving node and the reference receiver are independently and identically distributed (i.i.d.). Using this i.i.d. property, we derive the Laplace transform of interference distribution that is  the main component of coverage probability analysis.  The new distance distributions and  coverage probability results are used to study several metrics related to classical and currently trending  problems of  wireless networks. 
 
 {\em System design insights.} Our analysis leads to three main \chb{insights and} design guidelines. First, we use the new distance distributions to study diversity loss {due to $\sir$ correlation   under selection combining scheme in a finite network. } 
 Our analysis reveals that neglecting    correlation in ${\sir}$ distribution significantly overestimates the performance of the selection combining scheme.  Second, using the coverage probability result, we characterize the ${\tt NSE}$ of the whole network. We observe three different trends for ${\tt NSE}$ by increasing number of simultaneously active links in our current setup:   i) ${\tt NSE}$ under \chb{uniform TX-selection}  policy decreases, ii) ${\tt NSE}$ under  \chb{$k$-closest TX-selection} policy with $k=1$ increases, and iii) there exists an optimal number of simultaneously active links that maximize ${\tt NSE}$ for \chb{$k$-closest TX-selection} policy with $k>1$. Third,  we use coverage probability result to characterize the throughput and
determine the optimal caching strategy that maximizes total
hit probability in {\em finite} wireless networks. Our analysis demonstrates
that the increasing number of active nodes has a conflicting
effect on the maximum hit probability and the throughput:
maximum hit probability decreases whereas throughput increases.
This shows that more  nodes can be simultaneously
activated as long as the hit probability remains acceptable.

   \begin{figure}[t!]
\centering{
        \includegraphics[width=.4\linewidth]{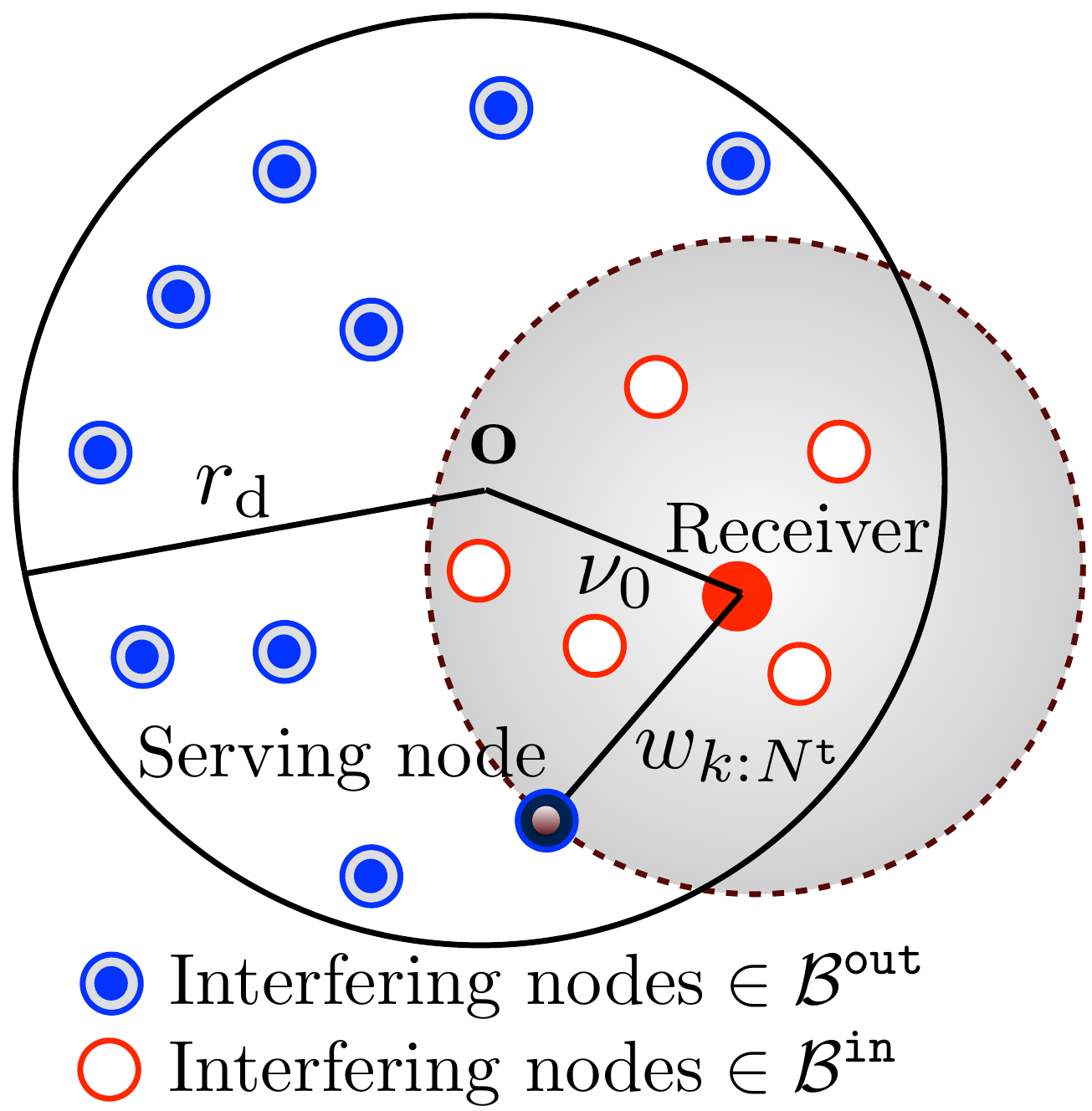}
              \caption{Illustration of the system model.}
                \label{Fig:sys_K}
                }
\end{figure}
\section{System Model} \label{sec:SysMod}
\subsection{System Setup and Key Assumptions}
We model the locations of   transmitting nodes as a uniform-BPP, where a fixed  number of nodes  are i.i.d.  in a finite region ${\cal A}\subset \R^2$. 
\chb{For notational and expositional simplicity, we consider ${\cal A}={\bf b}({\bf o}, r_{\rm d})$, which is a common assumption in the literature~\cite{srinivasa2010distance,torrieri2012outage,venugopal2015interference,venugopal2015analysis,valenti2014direct}. Here ${\bf b}({\bf o}, r_{\rm d})$ denotes ball of radius $r_{\rm d}$ that is centered at the origin. We assume $N^{\rm t}$ transmitting nodes are independently and uniformly distributed in this ball.}
 Denoting the locations  of transmitting nodes by  $\{{\bf y}_i\} \equiv \Phi_{\rm t}$, the probability density function (PDF) of each element ${\bf y}_i$  is:
\begin{align} f({\bf y}_i)
=
\begin{cases}
  \frac{1}{\pi r_{\rm d}^2}&   \|{\bf y}_i\|\le r_{\rm d}\\
  0               &  \text{otherwise}
\end{cases}.
\end{align}
\chb{It is worth noting that with some work our theoretical results can be extended to the case of arbitrarily-shaped polygon by using the methodology developed in~\cite{khalid2013distance}. This is however not in the scope of this paper.}
{We further assume that $N^{\rm a}$ out of $N^{\rm t}$ transmitting (serving and interfering) nodes
simultaneously reuse the same resource block. The locations of
simultaneously active nodes is denoted by  $\Phi^{\rm a} \subset \Phi^{\rm t}$. }  For this setup,
 we first perform analysis on the  reference receiver at an arbitrary location ${
\bf x}_0$  in ${\bf b}(\bf {o}, r_d) \subset \R^2$.  {Since uniform-BPP \chb{in ${\bf b}({\bf o}, r_{\rm d})$ is rotation invariant around the origin}, we assume that  {\tt x-axis} is  aligned with the location of reference receiver such that ${\bf x_0}=(\nu_0, 0)$, where $\nu_0=\|{\bf x}_0\|\in [0,r_{\rm d}]$}. We then  specialize and extend   the analysis of an arbitrarily-located reference receiver  to the two cases:
 i) {\em central receiver}, where the  reference receiver is located at the center of ${\bf b}({\bf o}, r_{\rm d})$, and ii) {{\em random receiver}, where the reference receiver  is chosen uniformly at {\em random} amongst  set of receiving nodes which are  independently and uniformly distributed  in ${\bf b}({\bf o}, r_{\rm d})$}}.

\subsection{ TX-selection Policies  and  Propagation Model}
We evaluate the network performance under two generic \chb{TX-selection} policies:
 \begin{enumerate}
 \item {\em uniform TX-selection policy}, where the serving node is  chosen uniformly at random from  $N^{\rm t}$  transmitting nodes. 
 \item {\em  $k$-closest TX-selection policy}, where  the serving node  is the $k^{th}$ closest node  out of $N^{\rm t}$   transmitting nodes to the reference receiver. 
 \end{enumerate}
These two policies  have various applications in ad-hoc and cellular networks. 
 For instance, the special case of $k=1$ can be used for modeling and analysis of downlink cellular network and generic \chb{$k$-closest TX-selection policy}  has  several applications \chb{to} the  performance evaluations of  emerging \chb{paradigms} such as
cache-enabled networks~\cite{BlaszczyszynG14}.  More details on the applications of these two policies will be discussed in Section~\ref{Applications of BPP}.  To keep the setup simple, we assume that the  background noise is  negligible 
compared to the  interference and is hence ignored. 
Denoting the location of the serving node with ${\bf y_\ell}$, the $\sir$ experienced by the reference receiver located at ${\bf x}_0$ is:
 \begin{equation}\label{eq: sir}
 \sir={\frac{   h_\ell \|{{\bf x}_0+\bf y}_\ell \|^{-\alpha}}{\sum_{{\bf y}_i\in \Phi_{\rm a}\setminus {\bf y}_\ell}  h_i\|{\bf x}_0+{\bf y}_i\|^{-\alpha}}},
  \end{equation}
 where $h_i\sim \exp(1)$ and $\|.\|^{-\alpha}$ model Rayleigh fading and
power law path-loss, respectively. It is important to note that after fixing the location of serving node for each policy, the interfering nodes (located at ${\bf y}_i \in \Phi_{\rm a} \setminus {\bf y}_\ell \subset \Phi_{\rm t}$) are assumed to be chosen uniformly at random amongst  set of possible transmitting nodes, i.e., $\Phi_{\rm t}$. For a quick reference, we summarize the notation of this paper in Table \ref{table:notation BPP}.

\begin{remark}[Scale invariance of BPP network] \label{Rem: Scale invariance of BPP network} 
\chb{Since the serving and interfering nodes are chosen from the same point process, the locations of these nodes with respect to the origin get scaled with the same factor when we change $r_{\rm d}$. This implies that the $\sir$ at the reference receiver located at ${\bf x}_0=(\nu_0,0)$, where  $\nu_0=\kappa_0 \: r_{\rm d}$ and $\kappa_0 \in [0, 1]$, is independent of the choice of $r_{\rm d}$ for  a given $\kappa_0$. Therefore, without loss of generality, we normalize $r_{\rm d}$ to 1.}
\end{remark}

For this setup, we are interested in studying the network performance in terms of coverage probability, which is formally defined next.
\begin{definition}[Coverage probability]  It is defined as the probability that $\sir$  at the reference receiver exceeds the predefined threshold needed to establish a successful connection. Mathematically, it is:
$$\pc=\E[{\bf 1}\{\sir\ge \beta \}]=\P(\sir\ge \beta),$$
where $\beta$ is   the minimum $\sir$ required to establish a successful connection.
\label{def: coverage}
\end{definition}
\begin{table}
\centering{
\caption{Summary of notation}
\scalebox{.8}{%
\begin{tabular}{c|c}
  \hline
   \hline
  \textbf{Notation} & \textbf{Description}  \\
     \hline
  $\Phi_{\rm t}; N^{\rm t}$ & Uniform-BPP modeling the locations of transmitting nodes; number of transmitting nodes\\
  \hline
  $\Phi_{\rm a}\subseteq \Phi_{\rm t}; N^{\rm a}$ & Set of simultaneously active transmitting nodes; number of simultaneously active transmitting nodes\\
  \hline
 $\ncalB^{\tt in}$ ($\ncalB^{\tt out}$)& Set of simultaneously active transmitting nodes closer (farther) than serving node to the reference reciver\\
    \hline
 $h_{i}$;$\alpha$  &Channel power gain under Rayleigh fading where $h_{i} \sim \exp (1)$;  path loss exponent where $\alpha>2$\\
    \hline
   $\pc;\T;{\tt NSE}$ &Coverage probability; target $\sir$; network spectral efficiency\\
  \hline
   ${\tt P}_{\rm ref}^{(u)}$ $({\tt P}_{\rm ref}^{(k)})$ &Coverage probability  of an arbitrarily-located reference receiver under uniform TX-selection  ($k$-closest TX-selection) policy\\
    \hline
     ${\tt P}_{\rm cent}^{(u)}$ $({\tt P}_{\rm cent}^{(k)})$ &Coverage probability  of a central receiver under uniform TX-selection  ($k$-closest TX-selection) policy\\
    \hline
        ${\tt P}_{\rm rand}^{(u)}$ $({\tt P}_{\rm rand}^{(k)})$ &Coverage probability  of a \chb{random} receiver under uniform TX-selection  ($k$-closest TX-selection) policy\\
          \hline
   ${\tt NSE}^{(u)}$  $({\tt NSE}^{(k)} )$ & Network spectral efficiency of uniform TX-selection ($k$-closest TX-selection) policy\\ 
  \hline
${\tt P}_{R_j}$;$b_j$; ${\tt P}_{\rm hit} $ & Request probability; caching probability; total hit probability\\
  \hline
    \hline

\end{tabular} \label{table:notation BPP}
}
} 
\end{table}
\section{Coverage Probability Analysis}
\label{sec: Coverage Probability Analysis}
This is the first main technical section of the paper, where we evaluate the  network performance in terms of coverage probability. Before going into the detailed analysis, we first characterize the distribution of the distances from the reference receiver to the transmitting  nodes in the next subsection. {This will be a key intermediate result in the coverage analysis.}
\subsection{Relevant Distance Distributions in a BPP}
\label{subsec: On the Distribution of Distance in BPP}
As stated above, the distribution of distances from the interfering and serving nodes hold a key to the derivation of  the coverage probability. If the reference receiver is assumed to be located at the origin (i.e., central receiver),  it is easy to
infer that the sequence of distances from transmitting nodes 
to the reference receiver which is denoted by   $\{W_i=\|{\bf y}_i\|\}$ contains
i.i.d. elements with PDF and CDF given by~\cite{srinivasa2010distance}:
\begin{align}\label{Eq:PDF unf}
\text{PDF}: \quad f_{W_i}(w_i)=\frac{2 w_i}{r_{\rm d}^2}; \quad 0 \le w_i \le r_{\rm d}, \quad \quad
\text{CDF}: \quad F_{W_i}(w_i)=\frac{w_i^2}{r_{\rm d}^2}; \quad 0 \le w_i \le r_{\rm d}.
\end{align}
 However, the sequence of distances from an arbitrarily-located reference receiver, i.e., $\{W_i=\|{\bf x}_0+{\bf y}_i\|\}$,  are correlated due to common factor ${\bf x}_0$. This means that if we condition on  $\bf x_0$, the set of distances $\{W_i=\|{\bf x}_0+{\bf y}_i\|\}_{i=1:N^{\rm t}}$ becomes conditionally i.i.d.  The conditional  CDF and PDF of each element of \{$W_i\}_{i=1:N^{\rm t}}$ are  stated in the next two Lemmas.
\begin{lemma}  \label{lem conditional_cdf_w}The conditional CDF of $W_i$  for a given $\nu_0=\|\nbx_0\|$ is: $F_{W_i}(w_i|\nu_0)
$
\begin{align}\label{eq:conditional_CDF_w}
&=
\begin{cases}
    F^{}_{W_{i,1}}(w_i|\nu_0)=\frac{ w_i^2}{ r_{\rm d}^2},&  0\le w_i\le r_{\rm d}-\nu_0\\
  F^{}_{W_{i,2}}(w_i|\nu_0)=\frac{w_i^2}{\pi r_{\rm d}^2}(\theta^{*}-\frac{1}{2} \sin 2 \theta^*){+}\frac{1}{\pi}(\phi^*-\frac{1}{2} \sin 2 \phi ^*) ,              &  r_{\rm d}-\nu_0 < w_i \le r_{\rm d}+\nu_0
\end{cases},
\end{align}
where  { $\theta^*=\arccos \big(\frac{w_i^2+\nu_0^2-r_{\rm d}^2}{2 \nu_0 w_i}\big)$}, and {$\phi^*=\arccos\big(\frac{\nu_0^2+r_{\rm d}^2-w_i^2}{2 \nu_0 r_{\rm d} }\big)$}.
 \end{lemma}
 \begin{IEEEproof}
As noted already, the distances between the transmitting nodes and a reference receiver are independent of coordinates system. Here, we  assume that the reference receiver lies on the positive side of {\tt x-axis}, and hence conditioning on $\nu_0=\|\nbx_0\|$, instead of $\nbx_0$ suffices. 
The CDF of $F_{W_i}(w_i|\nu_0)$ can be derived by using the same geometric argument applied in \cite[Theorem 2.3.6]{mathai1999introduction}. For completeness the proof is provided in the Appendix \ref{App: Lemma CDF distance}. \end{IEEEproof}
 \begin{lemma}
 \label{lem:conditional_pdf_w}
 The conditional PDF of $W_i$ for a given $\nu_0$ is: 
\begin{align}\label{eq:conditional_pdf_w}
f_{W_i}(w_i|\nu_0)
=
&\begin{cases}
    f^{}_{W_{i,1}}(w_i|\nu_0)= \frac{2 w_i}{ r_{\rm d}^2},&  0\le w_i\le  r_{\rm d}-\nu_0\\
  f^{}_{W_{i,2}}(w_i|\nu_0)= \frac{2 w_i}{\pi r_{\rm d}^2}\arccos \big(\frac{w_i^2+\nu_0^2-r_{\rm d}^2}{2 \nu_0 w_i}\big) ,              &   r_{\rm d}-\nu_0 < w_i \le  r_{\rm d}+\nu_0
\end{cases}.
\end{align}
  \end{lemma}
\begin{IEEEproof}
  $f^{}_{W_i}(w_i|\nu_0)$ can be  derived by taking the  derivative of  $F_{W_i}(w_i|\nu_0)$ with respect to $w_i$, and using basic algebraic manipulations. It is to be noted that this PDF is also provided in \cite{khalid2013distance}.
\end{IEEEproof}
As discussed above the $N^{\rm t}$ elements of the sequences of distances $\{W_i=\|{\bf x}_0+{\bf y}_i\|\}_{i=1:N^{\rm t}}$ are conditionally i.i.d. with density functions characterized by Lemmas~\ref{lem conditional_cdf_w} and \ref{lem:conditional_pdf_w}. This i.i.d. property \chb{is useful in characterizing} the distributions of serving and interfering distances for \chb{the two TX-selection} policies. We first focus on the distributions of various distances for  \chb{uniform TX-selection} policy, where the serving distance is one the elements of $\{W_i\}_{i=1:N^{\rm t}}$ that is chosen uniformly at random. The random selection of serving distance infers that the density function of serving distance simply follows that of ${W}_i$ given by Lemma \ref{lem:conditional_pdf_w}.
 Denoting the serving distance by $R=\|{\bf x}_\ell+{\bf y}_i\|$, the  conditional PDF of serving distance corresponding to the \chb{uniform TX-selection} policy is:
\begin{equation}\label{eq: serving link uni}
f^{(u)}_R(r|\nu_0)=f_{W_i}(r|\nu_0).
\end{equation}
Similarly, the $N^{\rm a}-1$ elements of  interfering distances $\{U = \|{\bf x}_0 +
{\bf y}_i\|, i\neq\ell\}$ are chosen uniformly at random. Hence the elements of $\{U\}$ are conditionally i.i.d., where the PDF of each element is:
\begin{equation} \label{eq:dis interfere unif}
f_{U}(u|\nu_0)=f_{W_i}(u|\nu_0),
 \end{equation}
where subscript $i$ is dropped for notational simplicity. We now focus on \chb{$k$-closest TX-selection} policy, where the serving node is the $k^{th}$ closest node to the reference receiver. Therefore, it is required to ``order" the distances from transmitting nodes to the reference receiver to characterize the density functions of serving and interfering distances. We define an ordered set $\{w_{i:N^{\rm t}}\}_{i=1:N^{\rm t}}$ by sorting  the value of $w_i$-s in ascending order such that  $w_{1:N^{\rm t}}<w_{2:N^{\rm t}}<...< w_{N^{\rm t}:N^{\rm t}}.$ Using the conditionally i.i.d. property of $\{W_i\}$, the conditional PDF of serving distance $R=W_{k:N^{\rm t}}$ is:
\begin{align}\label{eq: serving k policy}
&f^{(k)}_R(r|\nu_0)= \begin{cases}
    f^{(k)}_{R,1}(r|\nu_0),&  0\le r\le r_{\rm d}-\nu_0\\
  f^{(k)}_{R,2}(r|\nu_0) ,              &  r_{\rm d}-\nu_0 < r \le r_{\rm d}+\nu_0
\end{cases}\\ \notag
\text{with  } f^{(k)}_{R,j}(r|\nu_0)=& \frac{N^{\rm t}!}{(k-1)!(N^{\rm t}-k)!} 
 {F_{W_{i,j}}(r|\nu_0)}^{k-1}  f_{W_{i,j}}(r| \nu_0) (1-F_{W_{i,j}}(r|\nu_0))^{N^{\rm t}-k}; \:\:j=\{1,2\}, 
\end{align}
where, $f^{(k)}_R(r|\nu_0)$ can be  obtained   from the PDF of the $k$ order statistics of the sequence of i.i.d. random variables $\{W_i\}_{i=1:N^{\rm t}}$  with sampling PDF $f^{}_{W_{i}}(w_i|\nu_0)$~\cite{ahsanullah2005order}. It is important to note that the
possible interfering nodes can lie at any place except the location of the serving node. This means that the $k^{th}$ closest transmitting node to the reference receiver is explicitly removed from the  interference field.  In order to  incorporate this in the analysis, we partition  the set of distances from active  transmitting nodes (located at ${\bf y}_i \in \Phi_{\rm a}$) to the \chb{random} receiver into three subsets $\{\ncalB^{\tt in},w_{k:N^{\rm t}}, \ncalB^{\tt out} \}$  such  that $\ncalB^{\tt in}$ and $\ncalB^{\tt out}$ represent the set of interfering nodes closer and farther to the reference receiver, respectively, compared to the serving node.  This setup is illustrated in \figref{Fig:sys_K}.
The  following  Lemma deals with conditional i.i.d. property of $U_{\tt in} \in \ncalB^{\tt in}$ and $U_{\tt out}\in \ncalB^{\tt out}$, and their density functions.

\begin{lemma}\label{lem: density function of interferer distance k}
  Under \chb{$k$-closest TX-selection} policy, the sequences of random variables $U_{\tt in} \in \ncalB^{\tt in}$   and $U_{\tt out} \in \ncalB^{\tt out}$ conditioned on  $r=w_{k:N^{\rm t}}$, and $\nu_0$ are  independent. Moreover, 
  
  i)  the elements in the sequence of random variables $U_{\tt in} \in \ncalB^{\tt in}$  conditioned on  $r=w_{k:N^{\rm t}}$, and $\nu_0$ are i.i.d., where the PDF of each element  is $f_{U_{\tt in}}(u_{\tt in}|\nu_0, r)$
  \begin{align} \label{eq: pdf uin}
  =\left\{
 \begin{array}{cc}
 \frac{f_{W_i}(u_{\tt in}|\nu_0)}{F_{W_i}(r|\nu_0)}=\left\{
 \begin{array}{cc}
 \frac{f_{W_{i,1}}(u_{\tt in}|\nu_0)}{F_{W_{i,1}}(r|\nu_0)}, &  0< r<w^{-}, 0<u_{\tt in}<r\\
 \frac{f_{W_{i,1}}(u_{\tt in}|\nu_0)}{F_{W_{i,2}}(r|\nu_0)}, &   w^{-}< r<w^{+}, 0<u_{\tt in}<w^{-}  \\
  \frac{f_{W_{i,2}}(u_{\tt in}|\nu_0)}{F_{W_{i,2}}(r|\nu_0)}, &   w^{-}< r<{w^+},   w^- <u_{\tt in}<r
 \end{array}\right.,&u_{\rm in}<r  \\
 0, & u_{\tt in}\geq r
 \end{array}\right.,
  \end{align}
  ii) the elements in the sequence of random variables $U_{\tt out} \in \ncalB^{\tt out}$  conditioned on  $r=w_{k:N^{\rm t}}$, and $\nu_0$ are i.i.d, where the PDF of each element is $f_{U_{\tt out}}(u_{\tt out}|\nu_0, r)$
\begin{align}
 =\left\{
 \begin{array}{cc}
 \frac{f_{W_i}(u_{\tt out}|\nu_0)}{1-F_{W_i}(r|\nu_0)}=\left\{
 \begin{array}{cc}
 \frac{f_{W_{i,1}}(u_{\tt out}|\nu_0)}{1-F_{W_{i,1}}(r|\nu_0)}, &  0< r<w^-, r<u_{\tt out}<w^- \\
 \frac{f_{W_{i,2}}(u_{\tt out}|\nu_0)}{1-F_{W_{i,1}}(r|\nu_0)}, &   0< r<w^-,  w^-<u_{\tt out}< w^+ \\
  \frac{f_{W_{i,2}}(u_{\tt out}|\nu_0)}{1-F_{W_{i,2}}(r|\nu_0)}, &   w^{-}< r<{w^+},   r <u_{\tt out}<w^+
 \end{array}\right., & u_{\tt out}> r\\
 0, & u_{\rm out}\le r
 \end{array}\right.,
  \end{align}
 with $w^{-}=r_{\rm d}-\nu_0$,  and $w^{+}=r_{\rm d}+\nu_0$, where $F_{W_i}(.|\nu_0)$, and $f_{W_i}(.|\nu_0)$  are given by Lemmas \ref{lem conditional_cdf_w}  and  \ref{lem:conditional_pdf_w}.
  \end{lemma}
  \begin{IEEEproof}
  See Appendix \ref{App: Lemma PDF inner and outer}.
\end{IEEEproof}
\chb{Please note that in} \cite{afshangMehrnazD2D1,afshang2015fundamentals},
 we proved  similar i.i.d. property for
the distribution of distances in Thomas cluster process.

\subsection{Laplace transform of interference}
In this subsection, we characterize  the Laplace transform of interference distributions for various choices of intended receiver and transmitter by using the density functions of distances  derived in the  previous subsection.  As will be evident in the sequel, the characterization
of Laplace transform of interference distribution is the key intermediate step in the coverage probability analysis. 
\subsubsection{Laplace transform of interference under  \chb{uniform TX-selection} policy}
The Laplace transform of interference under \chb{uniform TX-selection} policy is given in the next Lemma.
\begin{lemma} \label{lem: laplace unif} 
Under \chb{uniform TX-selection} policy,  the Laplace transform of interference distribution conditioned on the location of  reference receiver, i.e., $\nu_0=\|{\bf x}_0\|$, is $\ncalL_{\ncalI}^{(u)}(s|\nu_0)=$
\begin{align}\label{eq: laplace unif}
&\bigg( \frac{1}{ r_{\rm d}^2}{\cal C}(\alpha, s, r_{\rm d}-\nu_0)+
\int_{r_{\rm d} - \nu_0}^{r_{\rm d} + \nu_0} \frac{u}{1+s u^{-\alpha}}\frac{2}{\pi r_{\rm d}}\arccos \big(\frac{u^2+\nu_0^2-r_{\rm d}^2}{2 \nu_0 u}\big) \nrmd u \bigg)^{N^{\rm a}-1}, \\
&\text{with} \quad {{\cal C}(\alpha, s, x)}=x^2-x^2 \:{}_2F_1(1, \frac{2}{\alpha}, 1+\frac{2}{\alpha},{-x^{\alpha}/ s})),\label{eq: c(a,s,x)}
\end{align}
where ${}_2F_1(a,b;c;z) = 1+\sum_{k=1}^\infty \frac{z^k}{k!}\prod_{l=0}^{k-1} \frac{(a+l)(b+l)}{c+l}$.
\end{lemma}
\begin{IEEEproof} 
See Appendix \ref{App: Laplace unif reference}
\end{IEEEproof}
The Laplace transform of interference distribution given by Lemma \ref{lem: laplace unif}  reduces to the simple closed-form expression for the special case of central receiver.  This result is  presented in the next Corollary and can be readily proved by substituting $\nu_0=0$ in  Lemma \ref{lem: laplace unif}.
\begin{cor}Under \chb{uniform TX-selection} policy, the  Laplace transform of interference at the central receiver is:
\begin{align} \label{Eq:ap unif}
&\ncalL_{\ncalI}^{(u)}(s)=\left(\frac{1}{r_{\rm d}^2} {\cal C}(\alpha, s, r_{\rm d})\right)^{N^{\rm a}-1},
\end{align}
which for  $\alpha = 4$ simplifies to
\begin{align}
{\ncalL_{\ncalI}^{(u)}(s)=\left(1-\frac{\sqrt{s}}{r_{\rm d}^2} \arctan{\frac{r^2_{\rm d}}{\sqrt{s}}}\right)^{N^{\rm a}-1},}
\end{align}
where ${\cal C}(\alpha, s, r_{\rm d})$ is given by \eqref{eq: c(a,s,x)}.

\label{cor: unif central Laplace}
\end{cor}
 We will use this result to approximate the coverage probability of an arbitrarily-located reference receiver later in this section.
\subsubsection{Laplace transform of interference under {\chb{$k$-closest TX-selection} } policy}
As  stated in the previous subsection, the  potential interfering nodes can lie at any place except the location of the serving node (i.e., the $k^{th}$ closest node to the reference receiver).
We mathematically \chb{incorporated} this  by partitioning  the set of distances from interfering nodes to  the reference receiver into subsets: $\ncalB^{\tt in}$ and $\ncalB^{\tt out}$. In Lemma \ref{lem: density function of interferer distance k}, we formally showed that the sequence of distances $U_{\tt in}\in \ncalB^{\tt in}$ and $U_{\tt out}\in \ncalB^{\tt out}$  conditioned on the location of the serving node are  i.i.d. Now, using the PDF of distances presented in Lemma \ref{lem: density function of interferer distance k},  the ``exact'' expression of  the Laplace transform of interference distribution at an arbitrarily-located reference receiver is provided in the next Lemma.
\begin{lemma}\label{lem: k-closest} Under {\chb{$k$-closest TX-selection} policy}, the  Laplace transform of interference  conditioned on  the location of   reference receiver and the serving distance is:
\begin{align}
\ncalL_{\ncalI}^{(k)}(s|\nu_0,r)=\left\{
 \begin{array}{cc}
  \ncalA(s,r,\nu_0), & 0\le r \le w^{-}\\
  \ncalB(s,r,\nu_0), & w^{-}< r\le w^{+}
 \end{array}\right.,\quad \text{with}
  \end{align}
   \begin{multline}\notag
    {\cal A}(s,r,\nu_0)=\sum_{\ell=0}^{n^{\rm a}_{m}}  \xi(p,n^{\rm a}_{m}) \Bigg(\int_0^{r}\frac{1}{1+s   {u_{\tt in}}^{-\alpha}}  \frac{f_{W_{i,1}}(u_{\tt in}|\nu_0)}{F_{W_{i,1}}(r|\nu_0)} \nrmd u_{\tt in} \Bigg)^{\ell}\\ \times
\Bigg(\int_r^{w^-}\frac{1}{1+s   {u_{\tt out}}^{-\alpha}}  \frac{f_{W_{i,1}}(u_{\tt out}|\nu_0)}{1-F_{W_{i,1}}(r|\nu_0)} \nrmd u_{\tt out} 
+\int_{w^-}^{w^+}\frac{1}{1+s   {u_{\tt out}}^{-\alpha}} \frac{f_{W_{i,2}}(u_{\tt out}|\nu_0)}{1-F_{W_{i,1}}(r|\nu_0)} \nrmd u_{\tt out} \Bigg)^{N^{\rm a}-\ell-1}, \ \text{and}
 \end{multline}
  \begin{multline}\notag
 {\cal B}(s,r,\nu_0)= \sum_{\ell=0}^{n^{\rm a}_{m}}  \xi(p,n^{\rm a}_{m}) \Bigg(\int_0^{w^-}\frac{1}{1+s   {u_{\tt in}}^{-\alpha}}  \frac{f_{W_{i,1}}(u_{\tt in}|\nu_0)}{F_{W_{i,2}}(r|\nu_0)} \nrmd u_{\tt in}+\int_{w^-}^{r}\frac{1}{1+s   {u_{\tt in}}^{-\alpha}}  \frac{f_{W_{i,2}}(u_{\tt in}|\nu_0)}{F_{W_{i,2}}(r|\nu_0)} \nrmd u_{\tt in} \Bigg)^{\ell}\\ \times
\Bigg(\int_r^{w^+}\frac{1}{1+s   {u_{\tt out}}^{-\alpha}}  \frac{f_{W_{i,1}}(u_{\tt out}|\nu_0)}{1-F_{W_{i,2}}(r|\nu_0)} \nrmd u_{\tt out} 
 \Bigg)^{N^{\rm a}-\ell-1},
 \end{multline}
  where $w^{-}=r_{\rm d}-\nu_0$,   $w^{+}=r_{\rm d}+\nu_0$, $\xi(p,n^{\rm a}_{m})=\frac{p^\ell (1-p)^{N^{\rm a}-\ell-1} \binom{N^{\rm a}-1}{\ell}} {\sum_{\ell=0}^{n^{\rm a}_{m}}p^\ell (1-p)^{N^{\rm a}-\ell-1} \binom{N^{\rm a}-1}{\ell}}$, $p=\frac{k-1}{N^{\rm t}-1}$, and $n^{\rm a}_{m}=\min(k-1,N^{\rm a}-1)$.
\end{lemma}
\begin{IEEEproof}
See Appendix \ref{App: proof of lemma Laplace k comm}.
\end{IEEEproof}
 This result can be simplified further for the special case of the central receiver, where it reduces
to a  simple  expression presented in the next  Corollary.
\begin{cor}\label{cor: Laplace under k-closest}
The Laplace transform of interference distribution at the central receiver conditioned on the serving distance $r=w_{k:N^{\rm t}}$ is: 
\begin{align}\label{eq: Laplace k closest central}
\ncalL_{\ncalI}^{(k)}(s|r)=\sum_{\ell=0}^{n^{\rm a}_{m}}  \xi(p,n^{\rm a}_{m}) \left(\frac{{\cal C}(\alpha,s, r)}{r^2}\right)^{\ell} \left(\frac{{\cal C}(\alpha,s, r_{\rm d})-{\cal C}(\alpha,s, r)}{r_{\rm d}^2-r^2}\right)^{N^{\rm a}-\ell-1},
\end{align}
where ${\cal C}(\alpha,s, x)$ is given by \eqref{eq: c(a,s,x)}.
\end{cor}
\begin{IEEEproof} 
This result can be simply derived by substituting $\nu_0=0$ in  Lemma~\ref{lem: k-closest}, where the final expression is obtained by using \cite[eq (3.194.1)]{zwillinger2014table}.
\end{IEEEproof}
This simple expression will be used to characterize the performance of the central receiver and approximate the coverage probability of an arbitrarily-located reference receiver in the next subsection.
\subsection{Coverage Probability}
Using the Laplace transform of interference distributions derived in the previous subsection, we now derive the coverage
probability for the two \chb{TX-selection} polices.  We begin our discussion with the \chb{uniform TX-selection} policy.
\subsubsection{Coverage probability under  \chb{uniform TX-selection} policy} The coverage probability for  \chb{uniform TX-selection}  policy is presented in the next Theorem.
\begin{thm}[\chb{Uniform TX-selection} policy]  The coverage probability of the reference receiver located at distance $\nu_0=\|{\bf x}_0\|$ from origin is:
\begin{align}\label{eq: coverage ref unif}
{\tt P}_{\rm ref}^{(u)}(\nu_0)=\int_0^{r_{\rm d}-\nu_0}\ncalL_{\ncalI}^{(u)}(\T r^{\alpha}|\nu_0) f^{}_{W_{i,1}}(r|\nu_0) \nrmd r+ \int_{r_{\rm d}-\nu_0}^{r_{\rm d}+\nu_0}\ncalL_{\ncalI}^{(u)}(\T r^{\alpha}|\nu_0) f^{}_{W_{i,2}}(r|\nu_0) \nrmd r,
\end{align}
where $f^{}_{W_{i,1}}(.|\nu_0)$, $f^{}_{W_{i,2}}(.|\nu_0)$ are given by Lemma \ref{lem:conditional_pdf_w}, and $\ncalL_{\ncalI}^{(u)}(.|\nu_0)$ is given by Lemma~\ref{lem: laplace unif}.
\label{thm: coverage unif}
\end{thm}
\begin{IEEEproof}Using  the definition of coverage probability, we have
\begin{align*}
\P\big(h_\ell \ge \T r^\alpha{ \sum_{{\bf y}_i \in \Phi_{\rm a}\setminus {{\bf y}_{\ell}}} h_i\|\nby_i\|^{-\alpha}}\big)\stackrel{(a)}=&\E_R\Big[\exp\Big(-\T r^{\alpha} {\sum_{{\bf y}_i \in \Phi_{\rm a} \setminus {\bf y}_\ell}  h_i\|\nby_i\|^{-\alpha}} \Big)\big |R\Big]
\end{align*}
where $(a)$ follows from $h_\ell \sim \exp(1)$. From this step, the final result  can be obtained by using the definition of Laplace transform, followed by de-conditioning over serving distance $R$.
\end{IEEEproof}
From  Theorem \ref{thm: coverage unif}, two corollaries are in order. First, we specialize the result to the case of central receiver in  the next Corollary. This result can be easily proved by substituting $\nu_0=0$ in Theorem \ref{thm: coverage unif}, and using Laplace transform of interference distribution given by  Corollary \ref{cor: unif central Laplace}.
\begin{cor}[\chb{Uniform TX-selection} policy]\label{cor: coverage unif-central}
The coverage probability of the central receiver located at the origin is:
\begin{align}
{\tt P}_{\rm cent}^{(u)}=\int_0^{r_{\rm d}}\ncalL_{\ncalI}^{(u)}(\T r^{\alpha}) f^{}_{W_{i}}(r) \nrmd r,
\end{align}
where $f_{W_{i}}(.)$ and $\ncalL_{\ncalI}^{(u)}(.)$ are given by~\eqref{Eq:PDF unf} and~\eqref{Eq:ap unif}, respectively.
\end{cor}
Second, we generalize the result of Theorem \ref{thm: coverage unif}  to analyze the performance of a \chb{random} receiver, where the intended receiver is chosen uniformly at random.  This result can be simply obtained by de-conditioning the result of Theorem \ref{thm: coverage unif}  with respect to distance from a randomly chosen receiver to the origin. Now, recall that   the receiving nodes are independently and uniformly  distributed in ${\bf b}({\bf o}, \nu_0)$, and hence the PDF of distance from a \chb{random} receiver to the origin is~\cite{srinivasa2010distance}: 
\begin{align}\label{eq: fV0}
f_{V_0}(\nu_0)=\frac{2 \nu_0}{r_{\rm d}^2}.
\end{align}
  Using this PDF, the coverage probability of a \chb{random} receiver is  stated  next.
\begin{cor}[\chb{uniform TX-selection} policy]\label{cor: coverage unif-typical} The coverage probability of a \chb{random} receiver is:
\begin{align}\label{eq: coverage unif typical}
{\tt P}_{\rm rand}^{(u)}=\int_0^{r_{\rm d}}{\tt P}_{\rm ref}^{(u)}(\nu_0)f_{V_0}(\nu_0) \:{\rm d} {\nu_0},
\end{align}
where ${\tt P}_{\rm ref}^{(u)}(.)$ and $f_{V_0}(.)$ are given by \eqref{eq: coverage ref unif} and \eqref{eq: fV0}, respectively.
\end{cor}
\begin{figure}
 \begin{minipage}{.49\textwidth}
  \includegraphics[width=1\textwidth]{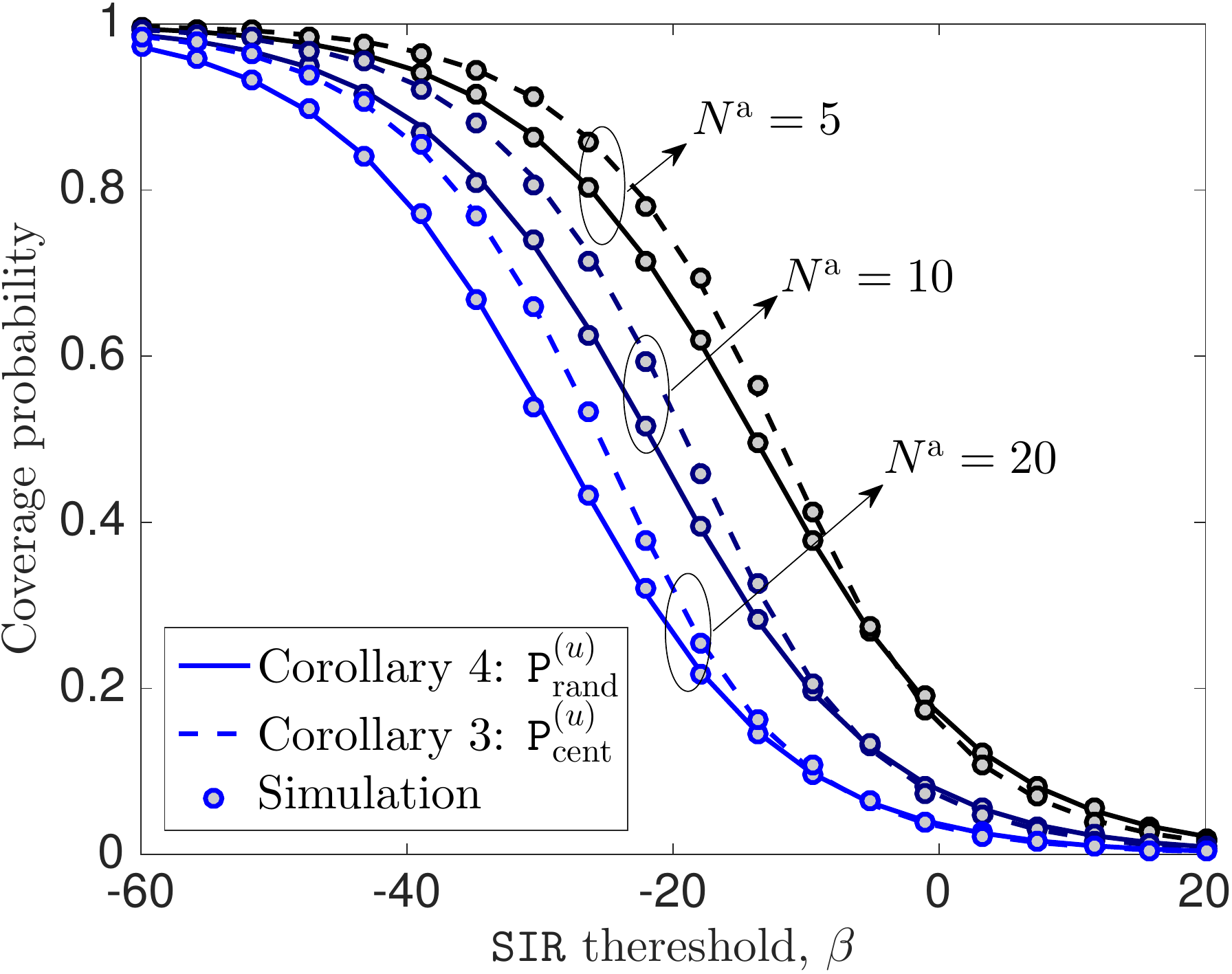}
              \caption{Coverage probability as a function of $\sir$ threshold ($\alpha=4$, and $r_{\rm d}=1$).}
                \label{Fig: Coverage typical central validation}
\end{minipage}%
\hfill
 \begin{minipage}{.49\textwidth}
  \includegraphics[width=1\textwidth]{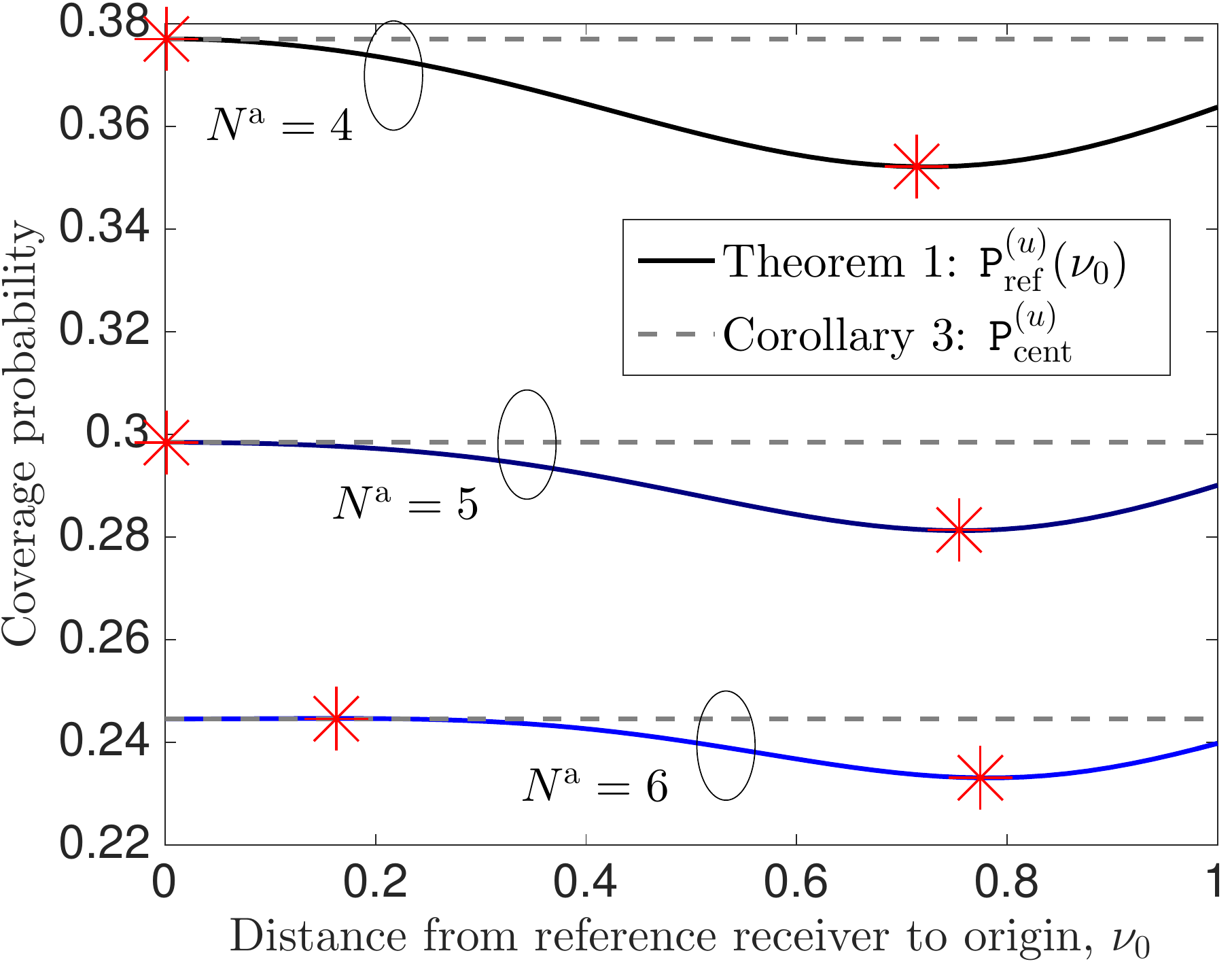}
              \caption{Coverage probability of reference receiver as a function of its distance to the origin ($\beta=-6$ dB, $\alpha=4$, and $r_{\rm d}=1$).}
\label{Fig:  Coverage reference unif v0 }
\end{minipage}
\end{figure}
\begin{remark}
\label{remark: approximation}
The coverage probability of the central receiver provides an approximation for that  of a \chb{random} receiver.
This is because putting the receiver at the center of ${\bf b}({\bf o}, r_{\rm d})$ has two conflicting  effects on the coverage probability: i) interfering link  distances decrease \chb{that increase interference power}, and ii) serving link distance decreases \chb{that increases received power of the desired signal}.
Depending upon the network parameters and \chb{TX-selection} policy, one of these effects (increasing interference/ received power)  dominates the other. Hence, the approximation provided by central receiver is not strictly a bound ( i.e., lower or upper bounds) of coverage probability of a \chb{random} receiver.

\end{remark}
We  plot the coverage probability of a \chb{random} receiver and central receiver  as  a function of $\sir$ threshold in \figref{Fig: Coverage typical central validation}.  The theoretical results of coverage probability under \chb{uniform TX-selection} policy  match perfectly  with simulation, thereby  validating the accuracy of the analysis. As formally stated in Remark \ref{remark: approximation} and evident from \figref{Fig: Coverage typical central validation}, the coverage probability of the central receiver given by  Corollary \ref{cor: coverage unif-central}  results in  approximation for that of a \chb{random} receiver. This observation emphasizes on the  importance of the accurate  \chb{random} receiver analysis.  In order to  understand  how the  performance of  the central receiver differs from that of an arbitrarily-located reference receiver,  \figref{Fig:  Coverage reference unif v0 } plots coverage probability of the reference receiver as a function of its distance from origin.  \chb{ In this and the rest of the plots, asterisk will be used to denote the extrema (maximum and minimum) of the curves.}
 As evident from \figref{Fig:  Coverage reference unif v0 },  the coverage probability  \chb{strongly} depends upon the location of the reference receiver. 
\begin{figure}
\center{
  \includegraphics[width=.49\textwidth]{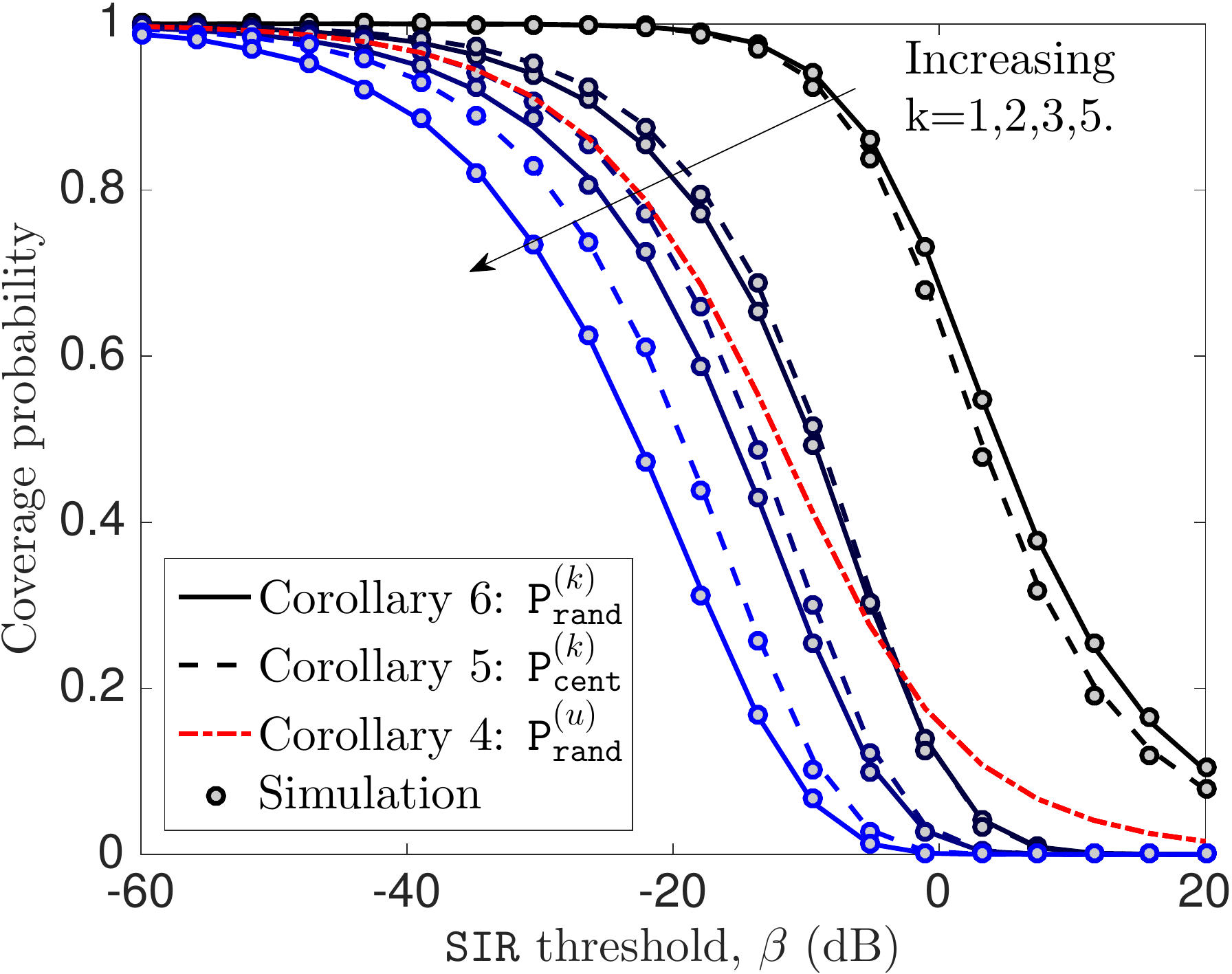}
              \caption{ Coverage probability as a function of $\sir$ threshold ($\alpha=4$, $r_{\rm d}=1$, and $N^{\rm a}=N^{\rm t}=5$)}
                \label{Fig: Coverage typical k-closet validation}
                }
\end{figure}
\begin{figure}
 \begin{minipage}{.49\textwidth}
  \includegraphics[width=1\textwidth]{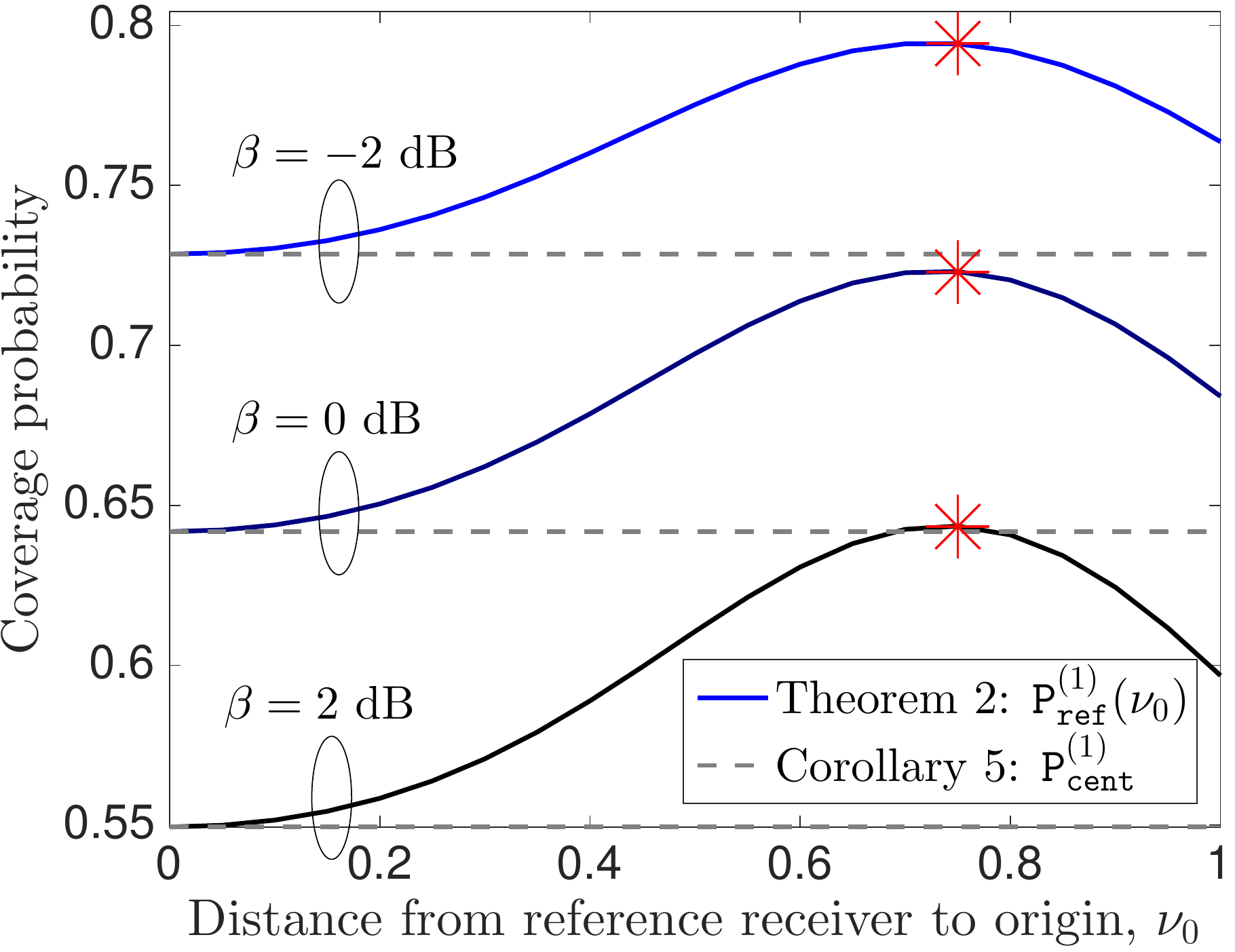}
              \caption{ Coverage probability of reference receiver as a function of its distance to the origin ($\alpha=4$, $r_{\rm d}=1$, and $N^{\rm a}=N^{\rm t}=5$)}
                \label{Fig: Coverage typical reference k=1}
\end{minipage}%
\hfill
 \begin{minipage}{.49\textwidth}
  \includegraphics[width=1\textwidth]{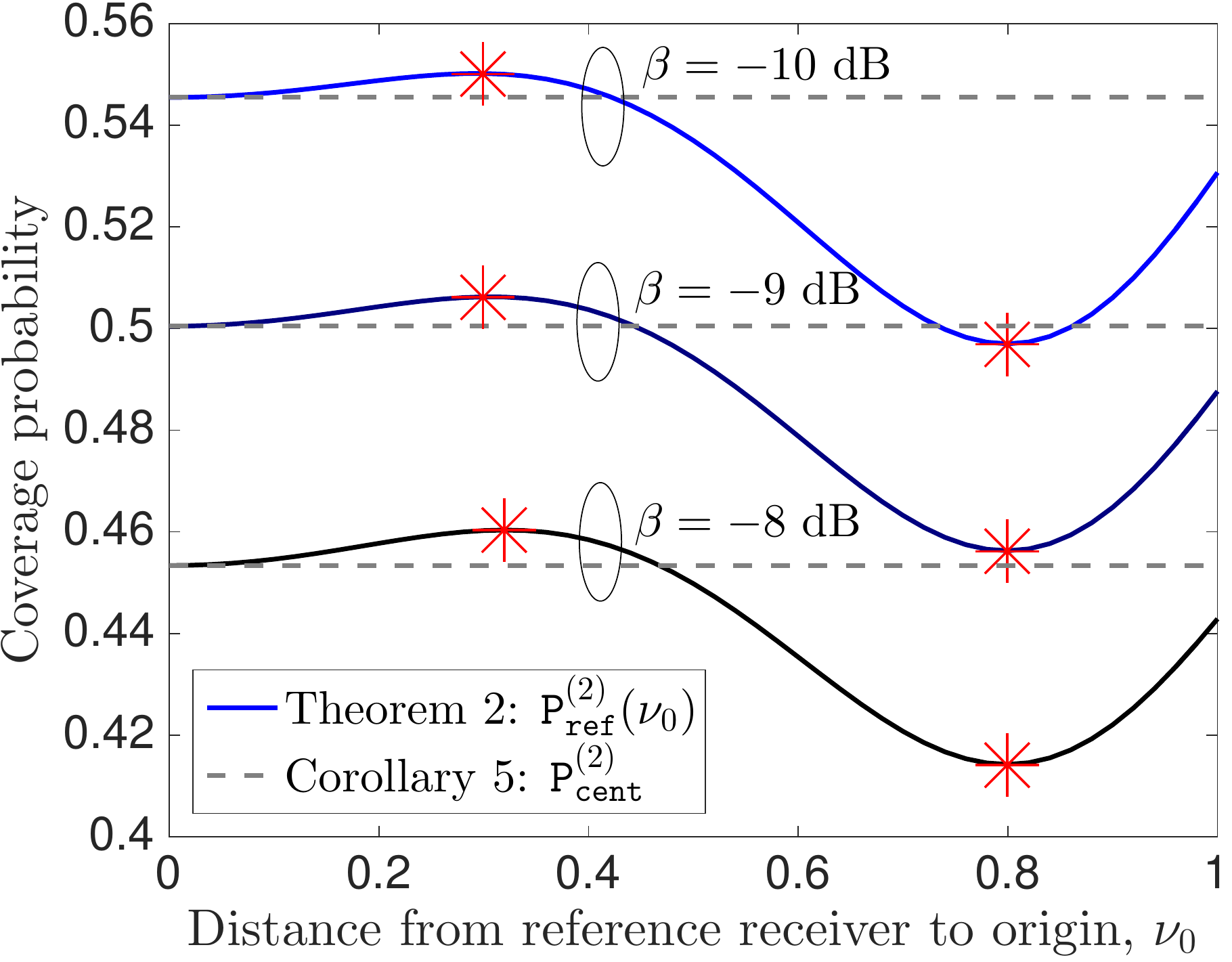}
              \caption{ Coverage probability of reference receiver as a function of its distance to the origin ($\alpha=4$, $r_{\rm d}=1$,  and $N^{\rm a}=N^{\rm t}=5$)}
\label{Fig:  Coverage reference  k=2}
\end{minipage}
\end{figure}

\subsubsection{Coverage probability under  \chb{$k$-closest TX-selection} policy}  We now use the Laplace transform of interference distribution derived in Lemma~\ref{lem: k-closest} to characterize the coverage probability of an arbitrarily-located reference receiver  under  \chb{$k$-closest TX-selection} policy in the next Theorem. \chb{As noted earlier, this result can be specialized to study coverage probability in a cellular network modeled as a finite network (such as in mmWave communications). It can also be used to study several emerging applications in wireless networks. Some examples will be provided in the next Section.}

\begin{thm}[\chb{$k$-closest TX-selection} policy]\label{them: coverage k-closest} Using the Laplace transform of interference given by Lemma~\ref{lem: k-closest},  the coverage probability of the reference receiver located at distance $\nu_0=\|{\bf x}_0\|$ from origin is:
\begin{align}\label{eq: coverage reference k comm}
{\tt P}_{\rm ref}^{(k)}(\nu_0)=\int_0^{r_{\rm d}-\nu_0} \ncalA(\T r^{\alpha},r,\nu_0)f^{(k)}_{R,1}(r|\nu_0)\nrmd r+ 
  \int_{r_{\rm d}-\nu_0}^{r_{\rm d}+\nu_0} \ncalB(\T r^{\alpha},r,\nu_0)f^{(k)}_{R,2}(r|\nu_0) \nrmd r,
\end{align}
where    $f^{(k)}_{R,j}(.|\nu_0)$ for $j\in\{1,2\}$  is given by \eqref{eq: serving k policy}.
\label{thm: k-comm}
\end{thm}
\begin{IEEEproof}
The proof follows on the same line as that of Theorem \ref{thm: coverage unif}.
\end{IEEEproof}
This result is specialized to the case of central receiver in  the next Corollary.
\begin{cor}[\chb{$k$-closest TX-selection} policy]\label{for: coverage k-closest central} The coverage probability of the central receiver for \chb{$k$-closest TX-selection} policy is:
\begin{align}
{\tt P}_{\rm cent}^{(k)}&=\int_0^{r_{\rm d}}\ncalL_{\ncalI}^{(k)}(\beta r^{\alpha}|r) f_R^{(k)}(r) {\rm d} r,\\
\text{with, } \quad f_R^{(k)}(r)&=\frac{N^{\rm t}!}{(k-1)!(N^{\rm t}-k)!}
 {F_{W_i}(r)}^{k-1}  f_{W_i}(r)(1-F_{W_i}(r))^{N^{\rm t}-k},
\end{align}
where  $f_{W_i}(.)$ and  $\ncalL_{\ncalI}^{(k)}(.|r)$   are given by \eqref{Eq:PDF unf}  and \eqref{eq: Laplace k closest central}, respectively.
\end{cor}
We further extend the result of Theorem \ref{thm: k-comm} to characterize the performance of a \chb{random} receiver. This result is presented in the next Corrolary.
\begin{cor}[\chb{$k$-closest TX-selection} policy]\label{for: coverage k-closest typical} The coverage probability of a \chb{random} receiver for \chb{$k$-closest TX-selection} policy is:
\begin{align}\label{eq: coverage k comm}
{\tt P}_{\rm rand}^{(k)}=\int_0^{r_{\rm d}}{\tt P}_{\rm ref}^{(k)}(\nu_0) f_{V_0}(\nu_0) {\rm d} \nu_0,
\end{align}
where $f_{V_0}(.)$ and ${\tt P}_{\rm ref}^{(k)}(\nu_0)$ are given by \eqref{eq: fV0} and \eqref{eq: coverage reference k comm}, respectively.
\end{cor}
\begin{IEEEproof}
This result can be simply obtained by taking expectation over the coverage probability of an arbitrarily-located reference receiver given by  \eqref{eq: coverage reference k comm} with respect to \chb{the distribution of $V_0$.}
\end{IEEEproof}
The coverage probabilities of the  \chb{random}  and central receivers as a function of $\sir$ threshold are presented in \figref{Fig: Coverage typical k-closet validation}. It is evident that  the analytical results  presented in  Corollaries \ref{for: coverage k-closest central} and \ref{for: coverage k-closest typical} match perfectly with simulation, which confirms the accuracy  of  the analyses.  Moreover, it can be   seen that the coverage probability of the central receiver leads to an approximation  for that of a \chb{random} receiver. The intuition  behind this approximation has been  stated in Remark \ref{remark: approximation}. We also plot  the coverage  probability  as a function of distance from reference receiver to the origin in Figs. \ref{Fig: Coverage typical reference k=1} and \ref{Fig:  Coverage reference  k=2}.  It can be observed that the coverage  probability strongly depends on the location of  the reference receiver, which again confirms the necessity of the  ``exact'' analysis for an  arbitrarily-located reference receiver.  This analysis can also be used  for characterizing the  performance of  the worst and best case receiver in a  finite wireless network.



\section{Applications and Discussion}\label{Applications of BPP}
This is the second main technical section of the paper, where we  use the distance distributions and coverage probability results presented in the above section to characterize  various classical and currently trending  problems related to  wireless networks. \chb{In particular, we study:}  i) diversity loss due to $\sir$ correlation under selection combining scheme in a finite  network, ii)  optimal number of simultaneously active links in a finite  network, and iii) optimal geographic caching in a finite network. 

\subsection{Diversity Loss Due to $\sir$ Correlation }
\chb{Diversity loss due to $\sir$ correlation in multi-antenna communication systems has been studied under different assumptions by modeling the system as infinite PPP in~\cite{haenggi2012diversity, Tanbourgi2014Correlation, Tanbourgi2015Heterogeneous, Tanbourgi2015Joint}. In this Section, we demonstrate how the distance distributions derived in subsection  \ref{subsec: On the Distribution of Distance in BPP} can be used to extend these analyses to finite network case. Due to space limitations, we will just study the performance of selection combining under interference correlation, which was presented for infinite PPP case in~ \cite{haenggi2012diversity}. As will be evident from the analysis, the same approach can be used to study other communications schemes, such as the automatic repeat request, where instead of combining signal in the spatial domain, it is combined over time domain. We consider the same setup as in the previous two sections with the only difference being that the reference receiver is now assumed to have $n>1$ antennas. The (Rayleigh) fading coefficients between transmitting nodes and the reference receiver are assumed to be independent across all links. Despite independent fading, the SIR observed across antennas at the reference receiver is correlated due to the common locations of the transmitting nodes. This correlation makes the analysis of coverage probability under diversity combining schemes challenging. } Before going in to the detailed \chb{coverage analysis}, we derive the probability of  joint occurrence of success event  \chb{at all $n$ antennas} which will  serve as \chb{the} basis \chb{for the coverage probability of selection combining under $\sir$ correlation.} Recall that the serving distance was denoted by $r$, and hence the joint success probability is defined as:
  \begin{equation}
 {\tt P}_{\rm joint}= \P(\cap_{j=1}^n \sir_j \ge \beta)= \P(\cap_{j=1}^n h_{\ell,j}>\beta r^{\alpha} {\cal I}_j),
   \end{equation}
 where  $\sir_j$ is the $\sir$ observed at the $j^{th}$ antenna,  ${\cal I}_j=\sum_{{\bf y}_i\in \Phi_{\rm a}\setminus {\bf y}_\ell}  h_{i,j} \|{\bf x}_0+{\bf y}_i\|^{-\alpha}$ and $h_i$ and  $h_{i,j}$ are exponential random variables with unit mean (modeling Rayleigh fading). We  study the impact of  correlation on \chb{the coverage probability under selection combining} for the two TX-selection policies next.
  \subsubsection{Uniform TX-selection policy} Using the density functions of distances presented in Lemma~\ref{lem:conditional_pdf_w}, the joint success probability for  \chb{uniform TX-selection} policy is stated in the next Lemma.
 \begin{lemma}\label{lem: joint success uniform}
 The joint success probability of the reference receiver for \ \chb{uniform TX-selection} policy is ${\tt P}_{\rm joint}^{(u)}(\nu_0,n)=$
 \begin{align}\label{eq: joint success unif}
&\int_0^{r_{\rm d}-\nu_0}\ncalL_{\ncalI_n}^{(u)}(\T r^{\alpha}|\nu_0) f^{}_{W_{i,1}}(r|\nu_0) \nrmd r+ \int_{r_{\rm d}-\nu_0}^{r_{\rm d}+\nu_0}\ncalL_{\ncalI_n}^{(u)}(\T r^{\alpha}|\nu_0) f^{}_{W_{i,2}}(r|\nu_0) \nrmd r,\ \text{with} \\\notag
&\ncalL_{\ncalI_n}^{(u)}(s|\nu_0)=\frac{1}{r_{\rm d}^2}\bigg({\cal D}(\alpha, s, r_{\rm d} - \nu_0,n)+
\int_{r_{\rm d} - \nu_0}^{r_{\rm d} + \nu_0}\left( \frac{1}{1+s u^{-\alpha}}\right)^n \frac{2 u}{\pi }\arccos \big(\frac{u^2+\nu_0^2-r_{\rm d}^2}{2 \nu_0 u}\big) \nrmd u \bigg)^{N^{\rm t}-1},
\end{align}
where ${{\cal D}(\alpha, s, x,n)}=\frac{2 x^2 (\frac{x^\alpha}{s})^n}{2+\alpha \: n} \:{}_2F_1(n, \frac{2}{\alpha}+n, 1+\frac{2}{\alpha}+n,{-x^{\alpha}/ s}))$, and
$f^{}_{W_{i,1}}(.|\nu_0)$, $f^{}_{W_{i,2}}(.|\nu_0)$ are given by Lemma \ref{lem:conditional_pdf_w}.
 \end{lemma}
 \begin{IEEEproof} See Appendix \ref{App: proof of Lemma joint success uniform}
  \end{IEEEproof}
 We now use the result of Lemma~\ref{lem: joint success uniform} to  study the effect of $\sir$ correlation on selection combining scheme where transmission is successful if $\max_{j \in \{1,2,..n\}} \sir_j \ge \beta$.  Using inclusion-exclusion principle~\cite{haenggi2012diversity}, the coverage probability under selection combining scheme \chb{can be equivalently expressed as:} ${\tt P}_{\rm SC}=$
 \begin{align}\label{eq: def selection combining}
 \P\big(\max_{m \in \{1,2,..n\}} \sir_m \ge \beta\big)= \P\big(\cup_{m=1}^n \sir_m \ge \beta \big)=\sum_{m=1}^n (-1)^{m+1} {n \choose m} \P(\cap_{j=1}^m \sir_j \ge \beta).
 \end{align}
  This definition can be readily used to characterize the coverage probability of a reference receiver for \chb{uniform TX-selection} policy under selection combining scheme. The  result is formally presented in the next Theorem.
  \begin{thm} [Uniform TX-selection policy] Using the joint success probability given by \eqref{eq: joint success unif}, the coverage probability of a reference receiver under selection combining scheme is:
   \begin{align}
  {\tt P}_{\rm SC-corr}^{(u)}(\nu_0,n)= \sum_{m=1}^n (-1)^{m+1} {n \choose m}{\tt P}_{\rm joint}^{(u)}(\nu_0,m),
   \end{align}
 where the probability of the same event, i.e., $\max_{m \in \{1,2,..n\}} \sir_m \ge \beta$  under the assumption of independent $\sir$ is:
    \begin{align}
{\tt P}_{\rm SC-ind}^{(u)}(\nu_0,n)= 1-\big(1-{\tt P}_{\rm ref}^{(u)}(\nu_0)\big)^n,
   \end{align}
   where ${\tt P}_{\rm ref}^{(u)}(\nu_0)$ is give by \eqref{eq: coverage ref unif}.
  \end{thm}
  
   \subsubsection{\chb{$k$-closest TX-selection} policy}  We now use the density functions of distances presented in Lemma~\ref{lem: density function of interferer distance k} to derive the joint  success probability of a reference receiver for \chb{$k$-closest TX-selection} policy. This result is formally stated in the next Lemma.
   \begin{lemma} \label{lem: joint success k comm}
   Using  the PDF of serving distances given by \eqref{eq: serving k policy},
    the joint success probability of a reference receiver for \chb{$k$-closest TX-selection} policy is ${\tt P}_{\rm joint}^{(k)}(\nu_0,n)=$
\begin{align}\label{eq: joint k comm}
\int_0^{r_{\rm d}-\nu_0} \ncalA_{\rm joint}(\T r^{\alpha},r,\nu_0,n)f^{(k)}_{R,1}(r|\nu_0)\nrmd r+ 
  \int_{r_{\rm d}-\nu_0}^{r_{\rm d}+\nu_0} \ncalB_{\rm joint}(\T r^{\alpha},r,\nu_0,n)f^{(k)}_{R,2}(r|\nu_0) \nrmd r, \: \text{where}
\end{align}
  \begin{multline}\notag
    {\cal A}_{\rm joint}(s,r,\nu_0,n)=\sum_{\ell=0}^{n^{\rm a}_{m}}  \xi(p,n^{\rm a}_{m}) \Bigg(\int_0^{r}\bigg(\frac{1}{1+s   {u_{\tt in}}^{-\alpha}}  \bigg)^n \frac{f_{W_{i,1}}(u_{\tt in}|\nu_0)}{F_{W_{i,1}}(r|\nu_0)} \nrmd u_{\tt in} \Bigg)^{\ell}\\
\Bigg(\int_r^{w^-}\bigg(\frac{1}{1+s   {u_{\tt out}}^{-\alpha}} \bigg)^n \frac{f_{W_{i,1}}(u_{\tt out}|\nu_0)}{1-F_{W_{i,1}}(r|\nu_0)} \nrmd u_{\tt out} 
+\int_{w^-}^{w^+}\bigg(\frac{1}{1+s   {u_{\tt out}}^{-\alpha}} \bigg)^n \frac{f_{W_{i,2}}(u_{\tt out}|\nu_0)}{1-F_{W_{i,1}}(r|\nu_0)} \nrmd u_{\tt out} \Bigg)^{N^{\rm a}-\ell-1},
 \end{multline}
  \begin{multline}\notag
 {\cal B}_{\rm joint}(s,r,\nu_0,n)= \sum_{\ell=0}^{n^{\rm a}_{m}}  \xi(p,n^{\rm a}_{m}) \Bigg(\int_0^{w^-}\bigg(\frac{1}{1+s   {u_{\tt in}}^{-\alpha}}  \bigg)^n\frac{f_{W_{i,1}}(u_{\tt in}|\nu_0)}{F_{W_{i,2}}(r|\nu_0)} \nrmd u_{\tt in}\\+\int_{w^-}^{r}\bigg(\frac{1}{1+s   {u_{\tt in}}^{-\alpha}} \bigg)^n \frac{f_{W_{i,2}}(u_{\tt in}|\nu_0)}{F_{W_{i,2}}(r|\nu_0)} \nrmd u_{\tt in} \Bigg)^{\ell}
\Bigg(\int_r^{w^+}\bigg(\frac{1}{1+s   {u_{\tt out}}^{-\alpha}}\bigg)^n\frac{f_{W_{i,1}}(u_{\tt out}|\nu_0)}{1-F_{W_{i,2}}(r|\nu_0)} \nrmd u_{\tt out} 
 \Bigg)^{N^{\rm a}-\ell-1}
 \end{multline}
  where $w^{-}=r_{\rm d}-\nu_0$,  and $w^{+}=r_{\rm d}+\nu_0$, $\xi(p,n^{\rm a}_{m})=\frac{p^\ell (1-p)^{N^{\rm a}-\ell-1} \binom{N^{\rm a}-1}{\ell}} {\sum_{\ell=0}^{n^{\rm a}_{m}}p^\ell (1-p)^{N^{\rm a}-\ell-1} \binom{N^{\rm a}-1}{\ell}}$, $p=\frac{k-1}{N^{\rm t}-1}$, $n^{\rm a}_{m}=\min(k-1,N^{\rm a}-1)$.
   \end{lemma}
   \begin{IEEEproof}
  The proof follows on the same line as that of Lemma~\ref{lem: joint success uniform} where    ${\cal A}_{\rm joint}(.)$ and     ${\cal B}_{\rm joint}(.)$ can be derived by using the same argument applied in the proof of Lemma~\ref{lem: k-closest}, and is hence skipped.
   \end{IEEEproof}
 Now using the result of Lemma~\ref{lem: joint success k comm},  the coverage probability  of a reference receiver for \chb{$k$-closest TX-selection} policy under selection combining scheme is presented next.
 \begin{figure}
 \begin{minipage}{.49\textwidth}
  \includegraphics[width=1\textwidth]{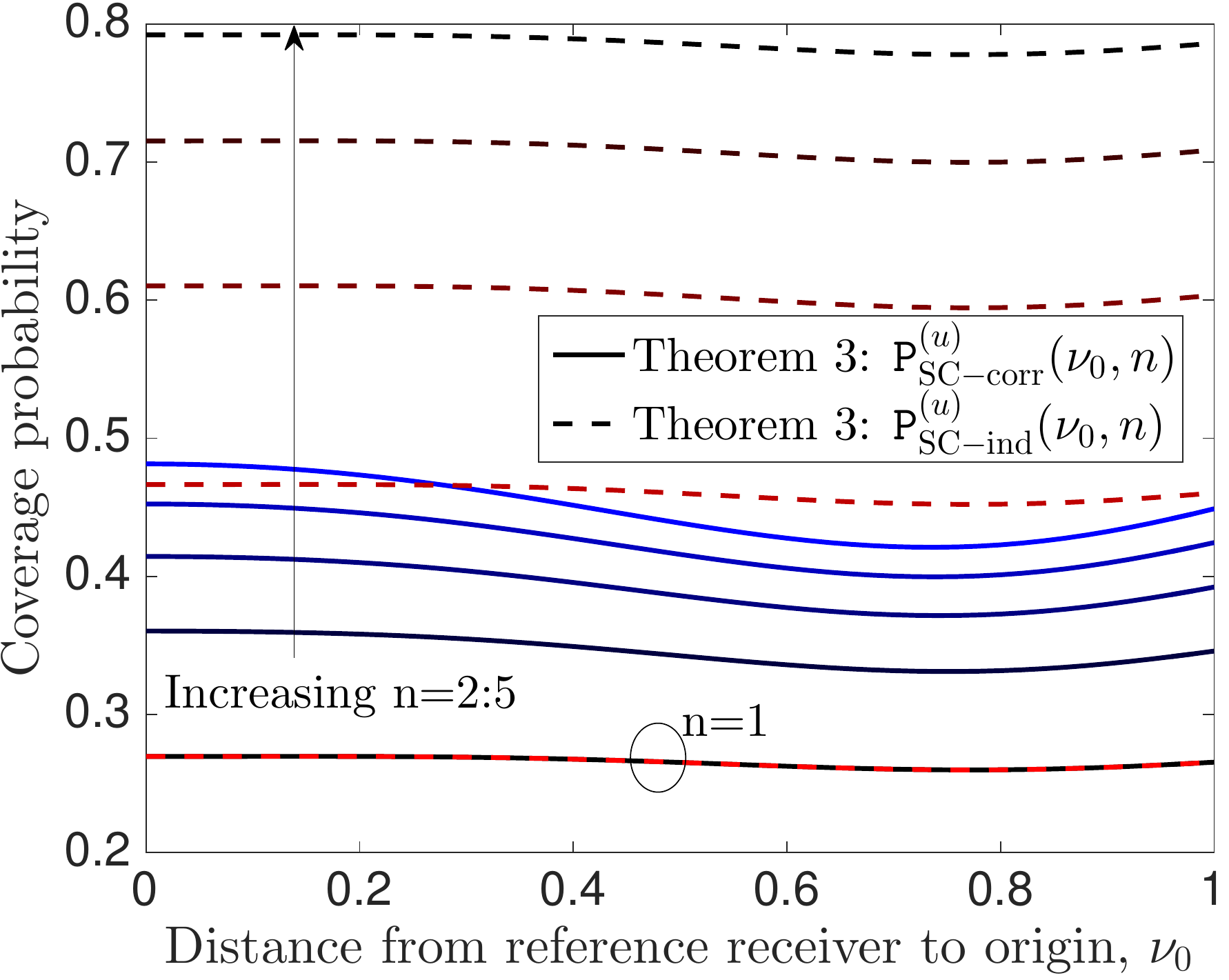}
              \caption{ Coverage probability of reference receiver as a function of its distance to the origin ($\alpha=4$, $r_{\rm d}=1$, $N^{\rm a}=5$, and $\beta=-5$ dB)}
                \label{Fig: Coverage ref unif SC}
\end{minipage}%
\hfill
 \begin{minipage}{.49\textwidth}
  \includegraphics[width=1\textwidth]{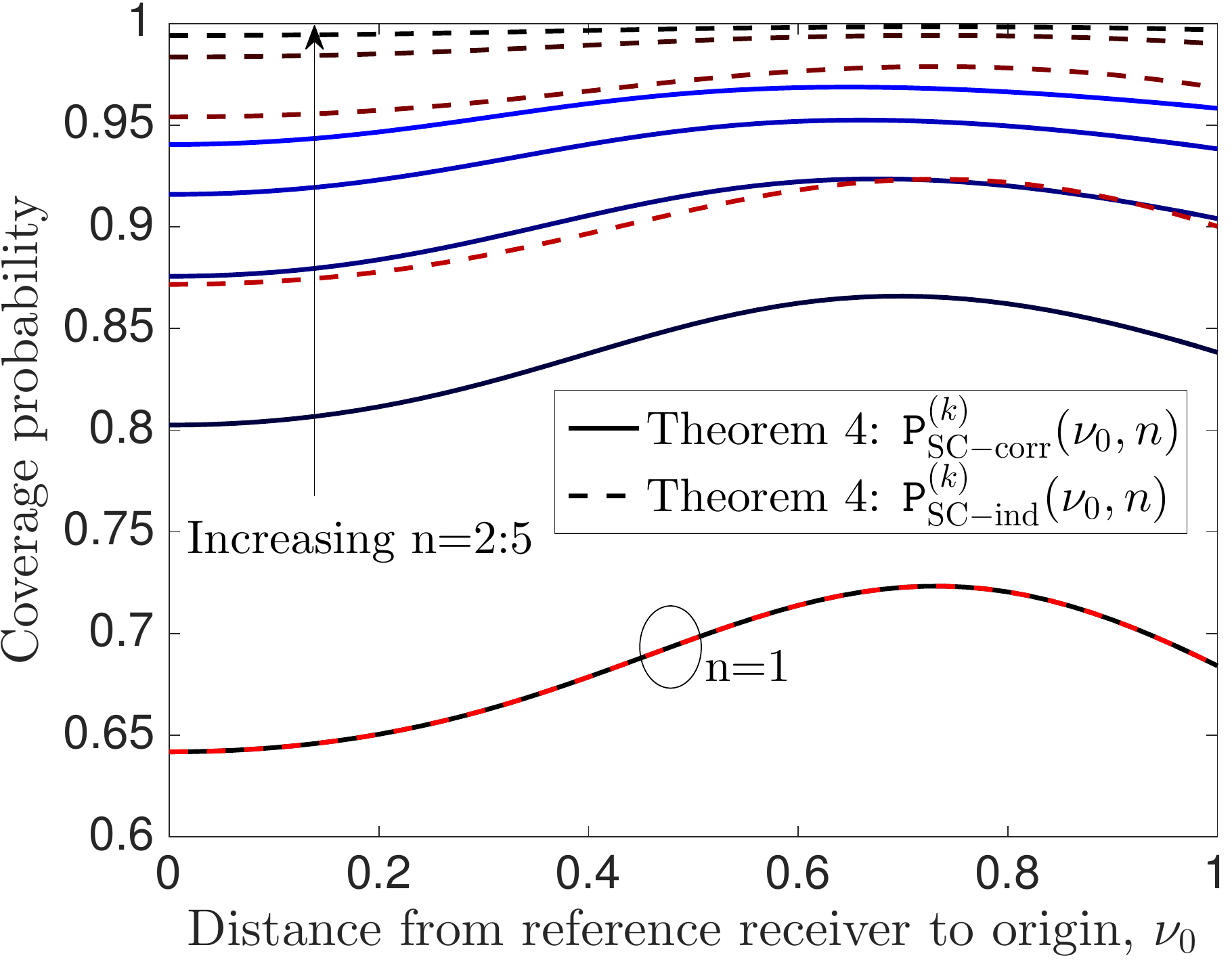}
              \caption{ Coverage probability of reference receiver as a function of its distance to the origin ($\alpha=4$, $r_{\rm d}=1$, $N^{\rm a}=N^{\rm t}=5$, and $\beta=0$ dB)}
\label{Fig:  Coverage ref k comm SC }
\end{minipage}
\end{figure}
 
   \begin{thm} [\chb{$k$-closest TX-selection} policy] Using the joint success probability given by \eqref{eq: joint k comm}, the coverage probability of a reference receiver under selection combining scheme is:
   \begin{align}
  {\tt P}_{\rm SC-corr}^{(k)}(\nu_0,n)= \sum_{m=1}^n (-1)^{m+1} {n \choose m}{\tt P}_{\rm joint}^{(k)}(\nu_0,m),
   \end{align}
 where the probability of the  event $\max_{m \in \{1,2,..n\}} \sir_m \ge \beta$  under the assumption of independent $\sir$ is:
    \begin{align}
{\tt P}_{\rm SC-ind}^{(k)}(\nu_0,n)= 1-\big(1-{\tt P}_{\rm ref}^{(k)}(\nu_0)\big)^n,
   \end{align}
   where ${\tt P}_{\rm ref}^{(k)}(\nu_0)$ is give by \eqref{eq: coverage reference k comm}.
  \end{thm}
 In Figs.~\ref{Fig: Coverage ref unif SC} and \ref{Fig:  Coverage ref k comm SC }, we plot the coverage probability under selection combining scheme for \chb{uniform TX-selection} and \chb{$k$-closest TX-selection} policies, respectively. It can be seen that \chb{the number of antennas at the receiver, $n$,}  the coverage probability of  both \chb{uniform TX-selection} and \chb{$k$-closest TX-selection} policies improve.
 \chb{ However, due to $\sir$ correlation, the actual gains are much smaller compared to the ones predicted under the independent $\sir$ assumption. This observation emphasizes the importance of the exact characterization of SIR correlation for the performance evaluation of diversity combining techniques. Note that this analysis for finite networks would not have been possible without the new distance distributions and coverage results derived in Section \ref{sec: Coverage Probability Analysis}.}

\subsection{Maximum Number of Simultaneously Active Links}
 In this subsection, we are interested in evaluating the {\em optimal} number of links that \chb{must} be simultaneously activated in a finite wireless network \chb{in the same time frequency resource block.}  In particular, we  study the  classical trade-off between  aggressive frequency reuse (i.e., more number of simultaneously active links on the same frequency band) and the resulting interference.  To study this trade-off, we  assume all transmitting nodes employ symbols from a Gaussian codebook for their transmissions, then, the minimum spectral efficiency (${\tt SE}$) of the links conditioned on the successful transmission (i.e., when  $\mathtt{SIR}> \beta$) is $\log_2(1+\beta)$. Hence, ${\tt SE}$ can be defined as:
 \begin{equation}
	\mathtt{SE} =\E[ \log_2(1+\beta)\,\mathbf{1}\{\mathtt{SIR}> \beta\}]= \log_2(1+\beta)\,\mathbb{P}\{\mathtt{SIR} > \beta\} \,\text{bits/s/Hz}.
	\label{eq:ShannonconditionalSE}
\end{equation}
Using this,  network spectral efficiency (${\tt NSE}$), i.e., total number
of bits transmitted per unit time per unit bandwidth \chb{across the whole network}, can be defined as: 
  \begin{equation}
	{\tt NSE} = N^{\rm a} \log_2(1+\beta)\,\mathbb{P}\{\mathtt{SIR} > \beta\} \,\text{bits/s/Hz}.
	\label{eq:TSE def}
\end{equation}
\chb{It should be noted that the ${\tt NSE}$ is simply a scaled version of the area spectral efficiency, which is a well known metric used typically in the analysis of infinite networks.}
The definition of ${\tt NSE}$ is specialized to our setup in the next Proposition.
 \begin{prop}[${\tt NSE}$] The  total number of bits transmitted per unit time per unit bandwidth for \chb{uniform TX-selection} policy is:
  \begin{equation}
	{\tt NSE}^{(u)} = N^{\rm a} \log_2(1+\beta)\,{\tt P}_{\rm rand}^{(u)}\ \text{bits/s/Hz}, 
	\label{eq:TSE unif comm }
\end{equation}
where ${\tt P}_{\rm rand}^{(u)}$ is given by \eqref{eq: coverage unif typical},
and under \chb{$k$-closest TX-selection} policy ${\tt NSE}$ is:
 \begin{equation}
	{\tt NSE}^{(k)} = N^{\rm a} \log_2(1+\beta)\,{\tt P}_{\rm rand}^{(k)}\ \text{bits/s/Hz}, 
	\label{eq:TSE k comm }
\end{equation}
where  ${\tt P}_{\rm rand}^{(k)}$ is given by \eqref{eq: coverage k comm}.
 \end{prop}
 
   \begin{figure}
   \begin{minipage}{.49\textwidth}
   \includegraphics[width=1\linewidth]{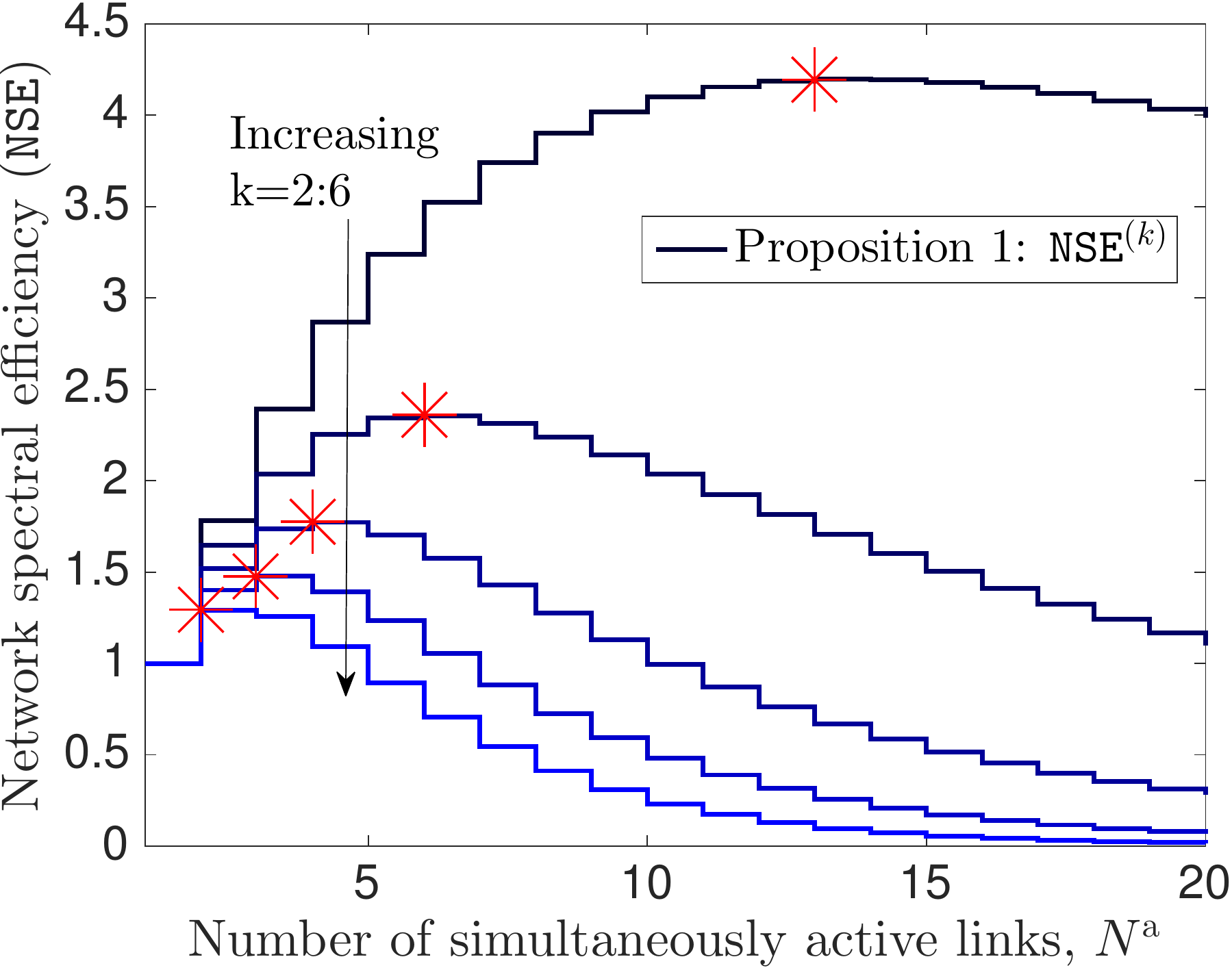}
              \caption{{${\tt NSE}$ as a function of number of active links ($r_{\rm d}=1$, $\beta=0$ dB, $\alpha=4$, and $N^{\tt t}=20$)} }
                \label{Fig: ASE vs number of active k}
\end{minipage}
\hfill
  \begin{minipage}{.49\textwidth}
   \includegraphics[width=1\linewidth]{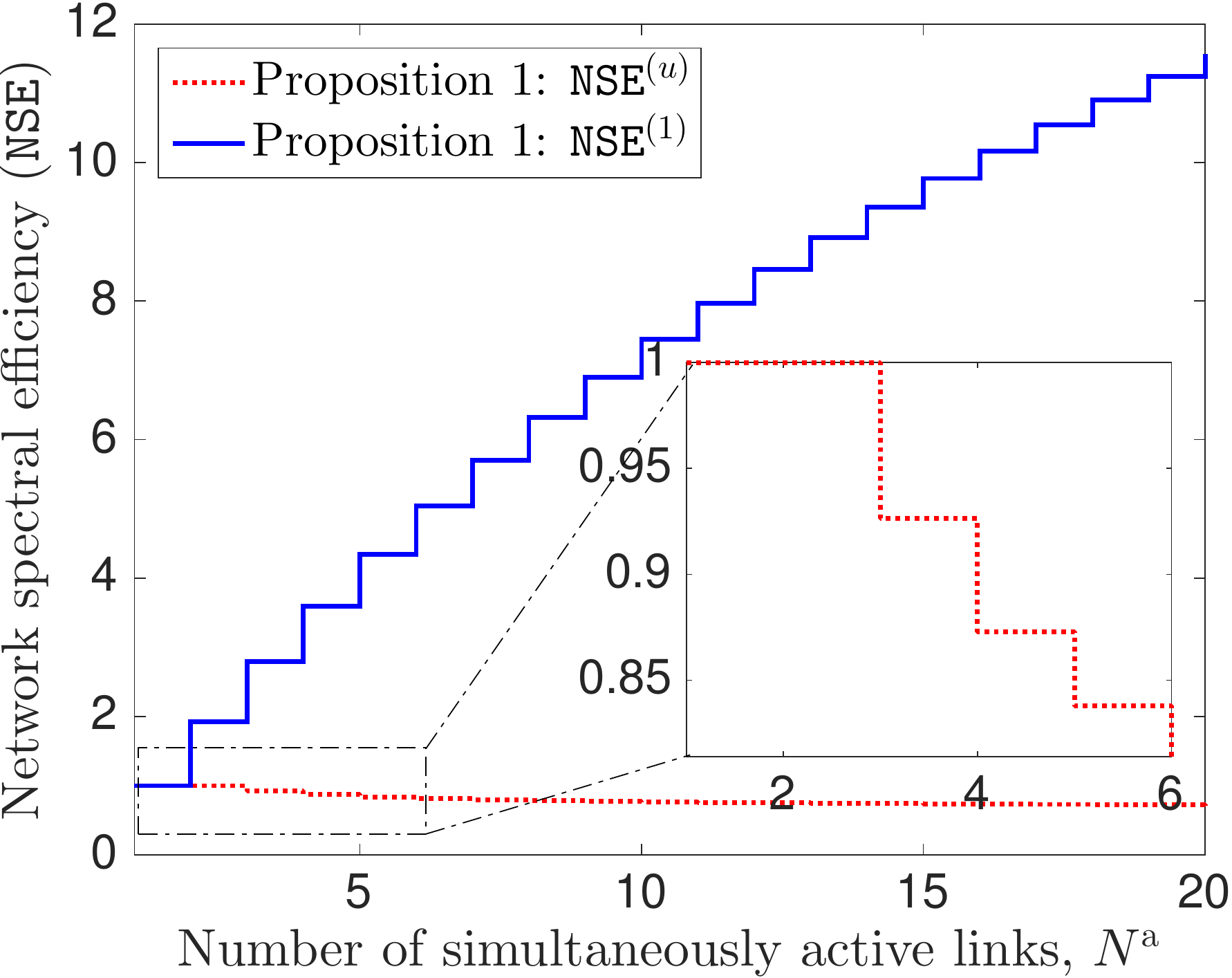}
              \caption{ ${\tt NSE}$ as a function of number of active links ($r_{\rm d}=1$, $\beta=0$ dB, $\alpha=4$, and $N^{\tt t}=20$) }
                \label{Fig: ASE vs number of active k1 unif}
\end{minipage}
\end{figure} 

In \figref{Fig: ASE vs number of active k}, we present ${\tt NSE}$ as  a function of   number of simultaneously active links  for different \chb{values} of $k$. Interestingly, it can be seen that when the distance between \chb{random} receiver and the  serving node  decreases, i.e., the value of $k$ is reduced,   the ``optimal number of simultaneously active links'' increases.
We also compare  ${\tt NSE}$ of \chb{uniform TX-selection} and \chb{$k$-closest TX-selection} ($k$=1) policies in \figref{Fig: ASE vs number of active k1 unif}.  It can be observed that increasing number  of simultaneously active links has conflicting effect on ${\tt NSE}$ of these two policies: ${\tt NSE}$ for \chb{uniform TX-selection} policy decreases and ${\tt NSE}$ for  \chb{$k$-closest TX-selection} ($k=1$) policy increases.

\subsection{ Optimal   Geographic Caching in Finite Wireless Networks}

In this subsection, we demonstrate the applicability of the tools developed in this paper to the performance analysis of cache-enabled finite networks. For detailed motivation and discussions, please refer to the shorter version of this paper~\cite{afshang2016optimal1}, where we studied this setup for the central user case. We consider the same setup as the previous two sections, except that the transmitting nodes are now assumed to have a local cache in which they can store some popular files that may be of interest to the other users. The metric of interest in this study is the total hit probability, which is the probability that the receiver of interest finds its requested content in one of the nodes that is accessible from this receiver. The total hit probability in turn depends upon: (i) {\em request probability}, which is the probability with which a particular content is requested by the reference receiver (assumed to be known {\em a priori}), (ii) {\em caching probability}, which is the probability with which a content is cached at a transmitting node (it depends upon the {\em caching strategy}), and (iii) coverage probability that guarantees the minimum SIR for successful reception~\cite{BlaszczyszynG14}.

\chb{Our goal is to maximize the total hit probability. To perform this analysis, we} consider a finite  library of popular contents ${\cal C}=\{{ c}_1,c_2,..., c_{\cal J}\}$ with size ${\cal J},$ where $c_j$ denotes the {$j^{th}$} most popular content. For simplicity of exposition, we  assume that all the contents have  same size, which are normalized to one. We further assume that the transmitting nodes  are equipped with cache storage of size $m_{\rm c}$ and hence each node can  store at most $m_{\rm c}$ popular contents. Now, denoting the specific set of contents at  a generic  transmitting node {by} $\Omega$, the caching probability  is:
    $b_{j}=\P(c_j\in \Omega), $
  where $b_{j}$ denotes the probability that  the content $c_j$ is stored at a given transmitting node.
To model the content popularity in this system, we use  Zipf's distribution due to its practical relevance \cite{cha2007tube}. Hence,  the request probability for file $c_{j}$ is:
\begin{align}\label{eq: zipf}
{\tt P}_{R_j}=\frac{j^{-\gamma}}{\sum_{i=1}^{\cal J} i^{-\gamma}}; \quad 1 \le j \le {\cal J},
\end{align}
where $\gamma$ represents the parameter of Zipf's distribution.   Now,  the total hit probability can be mathematically expressed as:
 \begin{align}
{\tt P}_{\rm hit}=\sum_{j=1}^{{\cal J}}  {\tt P}_{R_j}  \sum_{k=1}^{N^{\rm t}} {\tt P}^{ (k)}_{{\rm rand}} (1-b_j)^{k-1} b_j,
\end{align}
where $(1-b_j)^{k-1} b_j $ indicates that the closest node with content of interest is the $k^{th}$ closest node to the \chb{random} receiver. \chb{Note that we considered random receiver in this case because by construction, we need a ``network-averaged'' metric.}

\begin{figure}[t!]
\centering{
        \includegraphics[width=.48\linewidth]{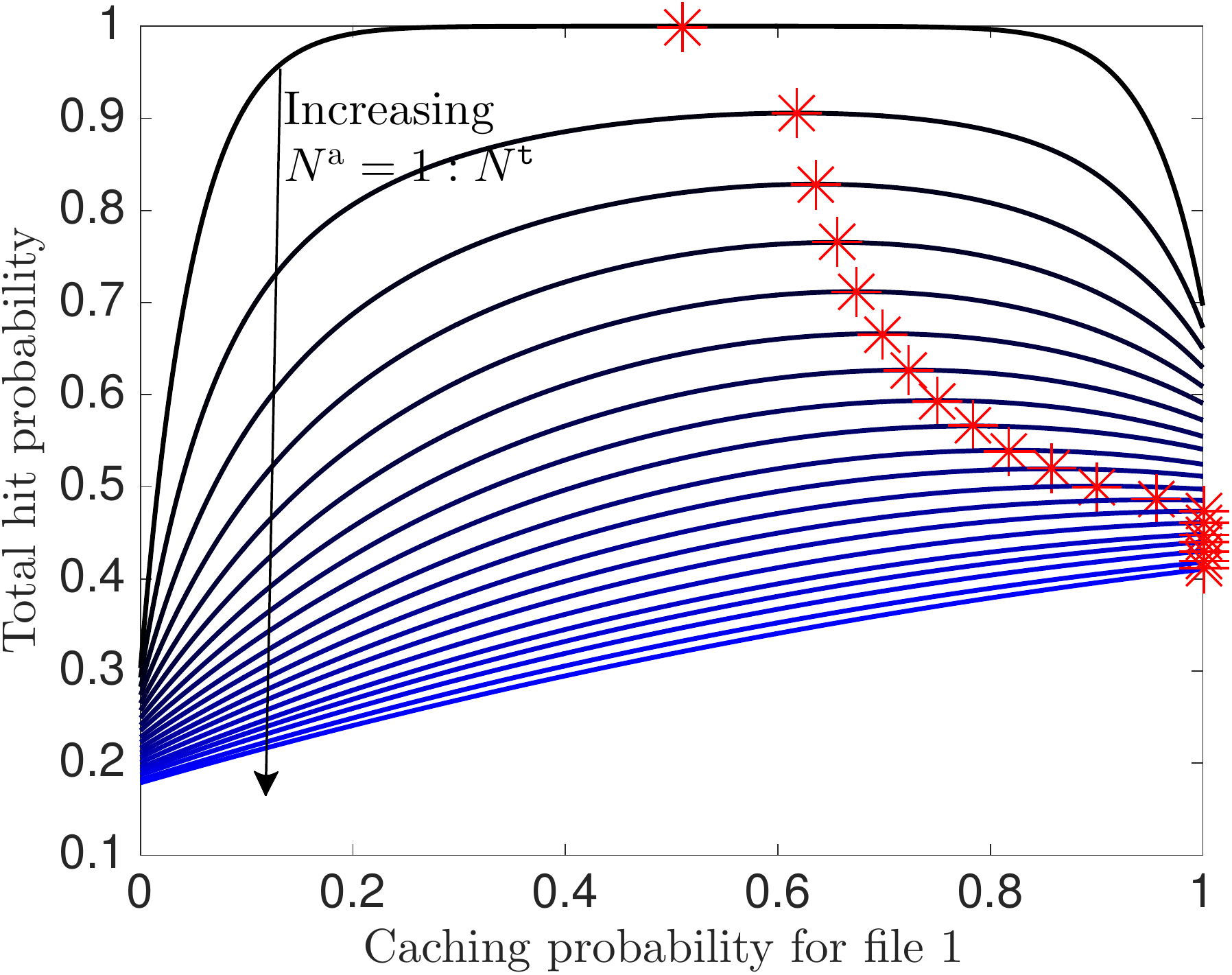}
              \caption{Total hit probability as  a function of caching probability for file 1, where asterisk shows optimal hit probability (${\cal J}=2$,  $m_{\rm c}=1$,  $\alpha=4$, $\beta=0$ dB, $\gamma=1.2$, and $N^{\tt t}=20$) }
                \label{Fig:optimal hit prob}
                }
\end{figure}

\begin{figure}
 \begin{minipage}{.49\textwidth}
 \includegraphics[width=\linewidth]{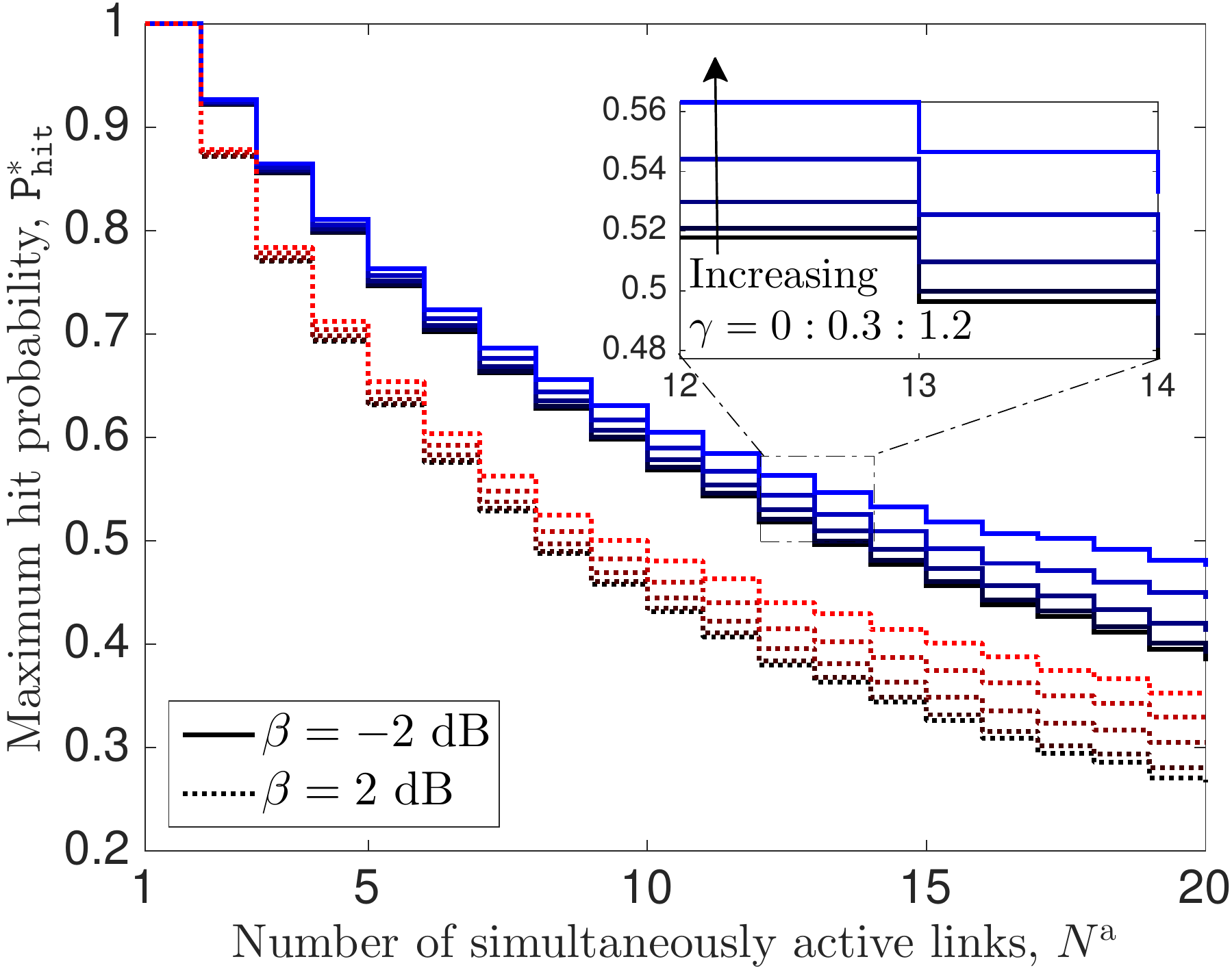}
              \caption{Maximum hit probability as a function of number of active devices ($r_{\rm d}=1$,  ${\cal J}=2$, $m_{\rm c}=1$, $\alpha=4$, and $N^{\tt t}=20$)}
                \label{Fig:optimal hit prob vs number of active}
\end{minipage}%
\hfill
 \begin{minipage}{.49\textwidth}
   \includegraphics[width=1\linewidth]{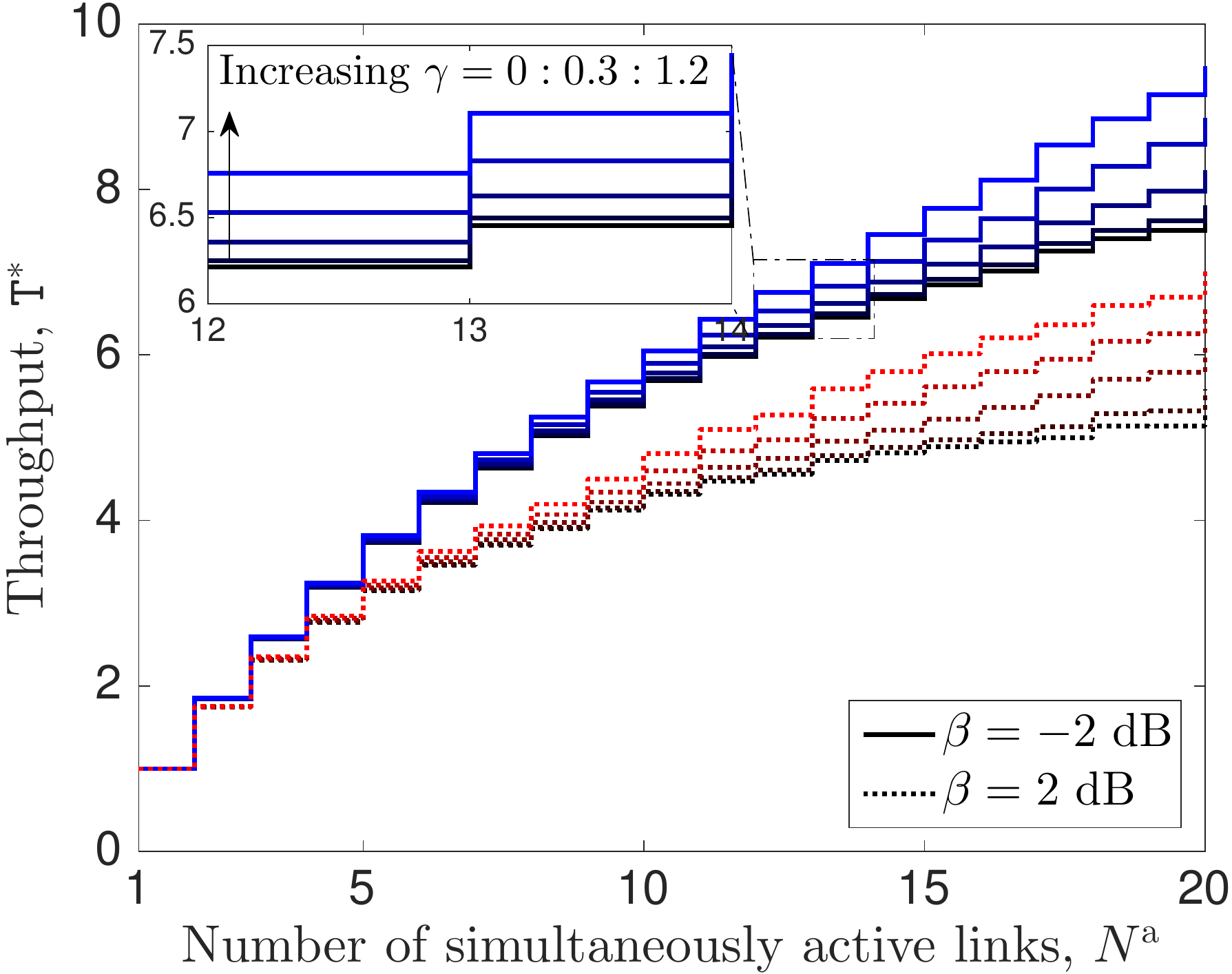}
              \caption{ Throughput as a function of number of active devices ($r_{\rm d}=1$,  ${\cal J}=2$, $m_{\rm c}=1$, $\alpha=4$, and $N^{\tt t}=20$) }
                \label{Fig:optimal ASE vs number of active}
\end{minipage}
\end{figure}

 In other words,
    the content of interest was not found at $(k-1)$ closet transmitting nodes to the \chb{random} receiver. On the same lines as \cite{BlaszczyszynG14}, the problem of optimal geographic caching in finite wireless networks can be formulated as:
       \begin{align}
&{\tt P}_{\rm hit}^{*}=   \max_{\{b_j\}} \sum_{j=1}^{{\cal J}}  {\tt P}_{R_j}  \sum_{k=1}^{N^{\rm t}} {\tt P}_{\rm rand}^{(k)} (1-b_j)^{k-1} b_j\\ \label{eq: constraint}
&\text{s.t.}\quad   \sum_{j=1}^{\cal J} b_j \le m_{\rm c}; \quad  0\le b_j \le 1\:\: \forall j,
\end{align}
where ${\tt P}_{\rm rand}^{(k)}$ is the new coverage probability result derived in Corollary~\ref{for: coverage k-closest typical}. The necessity and sufficiency of the constraints given by \eqref{eq: constraint} have already been discussed in \cite{BlaszczyszynG14}.
 For better understanding of this optimization problem, \figref{Fig:optimal hit prob} plots the total hit probability for the simple setup of ${\cal J}=2$, and $m_{\rm c}=1$.  It can be seen that by increasing the number of simultaneously active transmitting nodes, $N^{\rm a}$,
 the optimal caching probability for  the most popular content (i.e., $b_1$)  moves toward one. This is mainly because increasing $N^{\rm a}$ results in {higher} interference, {which in turn} decreases coverage probability.  For example when there is only one active node ({completely} orthogonal channel allocation), coverage probability is equal to {\em one} under interference limited regime.  Hence, it is optimal to {cache the two contents with the same probability, i.e.,} $b_1=b_2=0.5$. However, by increasing the number of active transmitting nodes, ${\tt P}_{\rm rand}^{(k)}$ for $k>1$ decreases significantly due to {increased} interference. {As demonstrated in \figref{Fig:optimal hit prob}, this makes it optimal to cache the more popular contents more often.}
 Though  channel orthogonalization is beneficial in terms of total hit probability, it is not {desirable} for the network throughput which {favors} having multiple active links as long as {the} resulting interference remains acceptable. In order to study the trade-off between {the} number of active links and the resulting interference, we  define throughput as: $ {\tt T}^{*}=N^{\rm a}\:{\tt P}_{\rm hit}^*
.$ Figs.~\ref{Fig:optimal hit prob vs number of active} and~\ref{Fig:optimal ASE vs number of active} plot maximum hit probability and  throughput versus number of active transmitting nodes, respectively. Interestingly, increasing the  number of active nodes has a conflicting effect on the maximum hit probability and the throughput: maximum hit probability decreases and throughput increases. 
 This implies that more nodes can be simultaneously activated if the total hit probability remains within acceptable limits.

\section{Conclusion}
In this paper, we developed a comprehensive framework for the performance analysis of  {\em finite} wireless networks. Modeling the locations of nodes as a uniform BPP, we considered two generic \chb{TX-selection} policies: i) \chb{uniform TX-selection} policy: the serving node is chosen uniformly at random amongst the set of transmitting nodes, and ii)  \chb{$k$-closest TX-selection} policy: the serving node is the $k^{th}$ closest node out of $N^{\rm t}$ transmitting nodes to the reference receiver.  For these two policies, we derived ``exact'' expressions of coverage probability corresponding to an arbitrarily-located reference receiver, using which we specialized and extended the analyses to \chb{several cases of interest.}  The new set of distance distributions and coverage probability results have numerous application in the performance analyses of various modern and classical wireless systems. \chb{For instance, \chb{$k$-closest TX-selection}  result can be easily specialized to study the performance of {\em finite} cellular networks. This case is becoming mainstream with the popularity of mmWave communications. We also discussed three possible applications of our new results.}
First, we investigated  the diversity loss due to $\sir$ correlation in a finite network. Second,  we obtained the optimal number of links that can be simultaneously activated to maximize ${\tt NSE}$. Third, we evaluated  optimal caching probability to maximize the total hit probability in cache-enabled finite networks.

This work has many extensions. From modeling perspective, it can be extended to an arbitrary shape (instead of a circle) where the node locations follow a more general distribution. Further, this framework can be extended to analyze Mat{\'e}rn cluster process where each finite network form one cluster. From system perspective, it can be used for the performance analysis of indoor communication and hotspots, where  the infinite PPP assumption may not be applicable. Finally, these tools can also be extended to study the performance of mmWave communication systems where the receiver of interest may experience interference from a finite number of nodes due to blocking.

\appendix
\subsection{Proof of Lemma \ref{lem conditional_cdf_w} }
\label{App: Lemma CDF distance}
The derivation of the CDF corresponding   to the distances between an arbitrarily-located reference receiver  to the transmitting nodes can be partitioned into two parts: a) $w_i \le r_{\rm d} - \nu_0$, and b) $w_i\ge r_{\rm d} - \nu_0$. First, if $w_i \le r_{\rm d} - \nu_0$, then $F^{}_{W_{i,1}}(w_i|\nu_0)$ simply is
$F^{}_{W_{i,1}}(w_i|\nu_0)=\frac{\pi w_i^2}{\pi r_{\rm d}^2}$.
Second, if $w_i> r_{\rm d}-\nu_0 $, then $F^{}_{W_{i,2}}(w_i|\nu_0)$ is the area of intersection of between the circles $x^2+y^2=r_{\rm d}^2$, and $(x-\nu_0)^2+y^2=w_i^2$ divided by $\pi r_{\rm d}^2$. These circles intersect  when $x=x^*= \frac{\nu_0^2+r_{\rm d}^2-w_i^2}{2 \nu_0}$. So, the CDF of $W_{i,2}$ can be written as:
\begin{align*}
F^{}_{W_{i,2}}(w_i|\nu_0)=\frac{1}{\pi r_{\rm d}^2} \Big[2 w_i \int_{\nu_0-w_i}^{x^*} \sqrt{(1-(({x-\nu_0})/{w_i})^2}{\rm d}x
+2 r_{\rm d}\int_{x^*}^{r_{\rm d}}\sqrt{1-(x/r_{\rm d})^2} {\rm d} x\Big].
\end{align*}
By changing variables $(\nu_0-x)/w_i=\cos \theta$ in the first integral  and {$x/r_{\rm d}= \cos \phi$} in the second integral, we have 
\begin{align*}
F^{}_{W_{i,2}}(w_i|\nu_0)&=\frac{1}{\pi r_{\rm d}^2}\left[2 w_i^2\int_0^{\theta^*}(1-\cos 2\theta){\rm d}\theta- 2 r_{\rm d}^2 \int_{\phi^*}^0 (1-\cos 2\phi){\rm d}\phi\right]\\
&=\frac{1}{\pi r_{\rm d}^2}\left[w_i^2(\theta^{*}-1/2 \sin 2 \theta^*) {+}r_{\rm d}^2(\phi^*-1/2 \sin 2 \phi ^*)\right],
\end{align*}
where  $\theta^*=\arccos \big(\frac{w_i^2+\nu_0^2-r_{\rm d}^2}{2 \nu_0 w_i}\big)$,  and $\phi^*=\arccos\big(\frac{\nu_0^2+r_{\rm d}^2-w_i^2}{2 \nu_0 r_{\rm d} }\big)$. 
\subsection{Proof of Lemma \ref{lem: density function of interferer distance k}}
\label{App: Lemma PDF inner and outer}

   \begin{figure}[t!]
\centering{
        \includegraphics[width=.95\linewidth]{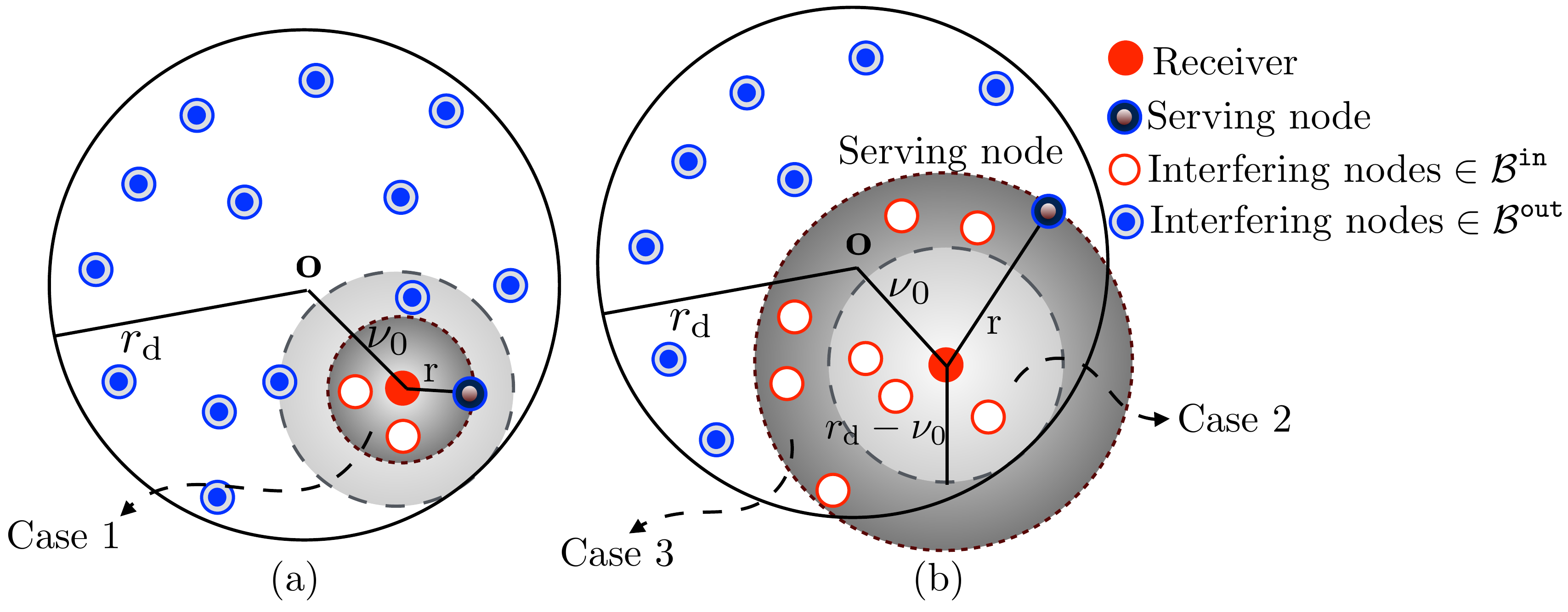}
              \caption{System model of \chb{$k$-closest TX-selection} policy wherein (a) $r <r_{\rm d}-\nu_0$, and in (b) $r >r_{\rm d}-\nu_0$.}
                \label{Fig:sys_Kproof}
                }
\end{figure}

The joint PDF of ``ordered" subset $\{w_{i:N^{\rm t}}\}_{i=1:N^{\rm t}}$ conditioned on $R=W_{k:N^{\rm t}}$ and $V_0$ is:
\begin{align}\notag 
f({{w_{1:N^{\rm t}}},...,{w_{N^{\rm t}:N^{\rm t}}}|w_{k:N^{\rm t}}},\nu_0)&=\frac{f({{w_{1:N^{\rm t}}},...,{w_{N^{\rm t}:N^{\rm t}}}}|\nu_0)}{f_{W_{k:N^{\rm t}}}(w_{k:N^{\rm t}}|\nu_0)}
\stackrel{(a)}{=}\frac{N^{t}! \prod_{i=1}^{N^{t}}f_{W_i}(w_{i:N^{\rm t}}|\nu_0)}{f_{W_{k:N^{\rm t}}}(w_{k:N^{\rm t}}|\nu_0)}\\\notag
&\stackrel{(b)}{=}\underbrace{(k-1)! \prod_{i=1}^{k-1} \frac{f_{W_i}(w_{i:N^{\rm t}}|\nu_0)}{F_{W_i}(w_{k:N}|\nu_0)} }_{\text{joint PDF of } \{W_{i:N^{\rm t}}\}_{i=1:k-1} }
 \underbrace{({N^{t}-k})! \prod_{i=k+1}^{N^{\rm t}} \frac{f_{W_i}(w_{i:N^{\rm t}}|\nu_0)}{1-F_{W_i}(w_{k:N}|\nu_0)}}_{\text{joint PDF of } \{W_{i:N^{\rm t}}\}_{i=k+1:N^{\rm t}} },\
\end{align}
where $f_{W_{k:N^{\rm t}}}(.|\nu_0)=f^{(k)}_R(.|\nu_0)$.  Here $(a)$ follows from definition of the joint PDF of order statistics \cite[eqn.~(2.10)]{ahsanullah2005order} with sampling PDF $f_{W_i}(.|\nu_0)$ given by \eqref{eq:conditional_pdf_w}, and (b) follows from  substituting the expression of  $f_{W_{k:N^{\rm t}}}(.|\nu_0)$ which is given by \eqref{eq: serving k policy}. The product of  joint PDFs in $(b)$   implies that the  $\{W_{i:N^{\rm t}}\}_{i=1:k}$  and $\{W_{i:N^{\rm t}}\}_{i=k+1:N^{\rm t}}$ conditioned on the  serving distance $R=W_{k:N}$ and $V_0$ are independent. The joint PDF of $\{W_{i:N^{\rm t}}\}_{i=1:k}$  conditioned on $R=W_{k:N^{\rm t}}$ and $V_0$ is:
\begin{align*}
f({{w_{1:N^{\rm t}}},...,{w_{k-1:N^{\rm t}}}|w_{k:N^{\rm t}}},\nu_0)=(k-1)! \prod_{i=1}^{k-1} \frac{f_{W_i}(w_{i:N^{\rm t}}|\nu_0)}{F_{W_i}(w_{k:N^{\rm t}}|\nu_0)}.
\end{align*}
Note that $(k-1)!$ \chb{gives} possible permutations of distances from ``ordered'' subset, $\{W_{i:N^{\rm t}}\}_{i=1:k-1}$, and  doesn't appear in the joint PDF of ``unordered'' subset  $\{W_{i}\}_{i=1:k-1}$, that is:
\begin{align*}
f({{w_{1}},...,{w_{k-1}}|w_{k:N^{\rm t}}},\nu_0)= \prod_{i=1}^{k-1} \frac{f_{W_i}(w_{i}|\nu_0)}{F_{W_i}(w_{k:N^{\rm t}}|\nu_0)};\:\: w_i\leq w_{k:N^{\rm t}},
\end{align*}
where the product of  same functional form of the joint PDF $f({{w_{1}},...,{w_{k-1}}|w_{k:N^{\rm t}}},\nu_0)$ infers that the subset of distances in ``unordered"  set $\{W_{i}\}_{i=1:k-1}$ are i.i.d. with PDF $\frac{f_{W_i}(w_{i}|\nu_0)}{F_{W_i}(w_{k:N^{\rm t}}|\nu_0)}$. Recall that after fixing the location of serving distance the  interfering nodes in $\ncalB^{\tt in}$ are chosen uniformly at random. This random selection of the interfering distances infers that the PDF of each element of distance in $\ncalB^{\tt in}$ is also $\frac{f_{W_i}(w_{i}|\nu_0)}{F_{W_i}(w_{k:N^{\rm t}}|\nu_0)}$.  Now, recall that  the distribution of serving distance \chb{has a piece-wise form}. We denote the distances in ``unordered" subset  $\ncalB^{\tt in}$ by $u_{\tt in}$, and substitute $w_{k:N^{\rm t}}$ with $r$. From \figref{Fig:sys_Kproof}, it can be seen that there are three cases for the distance of interfering nodes closer than serving node to the reference receiver.
\begin{itemize}
\item Case 1: the distances from interfering nodes that are closer than the  serving node to the reference receiver,   where the serving distance is smaller than $r_{\rm d}-\nu_0$, i.e.,
$u_{\tt in}<r <r_{\rm d}-\nu_0$.
\item Case 2: the distances from interfering nodes that are closer than the  serving node to the reference receiver  is less than  $r_{\rm d}-\nu_0$ and the serving distance  is greater than $r_{\rm d}-\nu_0$, i.e., $u_{\tt in}<r_{\rm d}-\nu_0 < r < r_{\rm d}+\nu_0$
\item Case 3: the distances from interfering nodes  that are closer than the  serving node to the reference receiver and  the serving distance  are greater than $r_{\rm d}-\nu_0$, i.e., $ r_{\rm d}-\nu_0< u_{\tt in}< r < r_{\rm d}+\nu_0$.
\end{itemize}
The appropriate pieces of PDF and CDF of $W_i$ (see Lemmas \ref{lem conditional_cdf_w}  and \ref{lem:conditional_pdf_w}) are  chosen according to the ranges of $r$ and $u_{\tt in}$ in each case. Thus, we have

 \begin{align*}
 \frac{f_{W_i}(u_{\tt in}|\nu_0)}{F_{W_i}(r|\nu_0)}=\left\{
 \begin{array}{cc}
\ \frac{f_{W_{i,1}}(u_{\tt in}|\nu_0)}{F_{W_{i,1}}(r|\nu_0)}, &  0< r<r_{\rm d}-\nu_0,  0<u_{\tt in}<r\\
\ \frac{f_{W_{i,1}}(u_{\tt in}|\nu_0)}{F_{W_{i,2}}(r|\nu_0)}, &   r_{\rm d}-\nu_0< r<r_{\rm d}+\nu_0, 0<u_{\tt in}<r_{\rm d}-\nu_0  \\
\  \frac{f_{W_{i,2}}(u_{\tt in}|\nu_0)}{F_{W_{i,2}}(r|\nu_0)}, &   r_{\rm d}-\nu_0< r<r_{\rm d}+\nu_0,   r_{\rm d}-\nu_0  <u_{\tt in}<r
 \end{array}\right.,
  \end{align*}
  where the CDF and PDF   of   ${W_{i,1}}$  and  ${W_{i,2}}$  are given by  Lemmas \ref{lem conditional_cdf_w}  and \ref{lem:conditional_pdf_w}. Similar arguments can be applied for the derivation of  $f_{U_{\rm out}}(.|\nu_0, r)$.

\subsection{Proof of Lemma \ref{lem: laplace unif}}
\label{App: Laplace unif reference}
The conditional  Laplace transform  of interference is:
\begin{align}\notag
\ncalL_{\ncalI}^{(u)}(s|\nu_0)&=\E\Big[\exp\Big(-s\sum_{{\bf y}_i \in \Phi_{\rm a}\setminus {\bf y}_\ell } h_i\|\nbx_0+\nby_i\|^{-\alpha}\Big)\Big]
=\E\Big[\prod_{{\bf y}_i \in \Phi_{\rm a}\setminus {\bf y}_\ell } \exp\Big(-s h_i\|\nbx_0+\nby_i\|^{-\alpha}\Big)\Big]\\ \notag
&\stackrel{(a)}=\E\Bigg[\prod_{{\bf y}_i \in \Phi_{\rm a}\setminus {\bf y}_\ell }\frac{1}{1+s \|\nbx_0+\nby_i\|^{-\alpha}}\Bigg] \stackrel{(b)}=\Bigg( \int_0^{r_{\rm d}+v_0}\frac{1}{1+s u^{-\alpha}} f_U(u|\nu_0) \nrmd u\Bigg)^{{N^{\rm a}}-1}\\
 &\stackrel{(c)}{=}\bigg( \frac{2}{ r_{\rm d}^2} \int_0^{r_{\rm d} - \nu_0} \frac{ u}{1+s u^{-\alpha}} \nrmd u+
\int_{r_{\rm d} - \nu_0}^{r_{\rm d} + \nu_0} \frac{u}{1+s u^{-\alpha}}\frac{2}{\pi r_{\rm d}}\arccos \big(\frac{u^2+\nu_0^2-r_{\rm d}^2}{2 \nu_0 u}\big) \nrmd u \bigg)^{N^{\rm a}-1},\notag
\end{align}
where $(a)$ follows from $h_i\sim \exp(1)$, $(b)$ follows from converting  Cartesian to polar coordinates  using density function of distance given by \eqref{eq:dis interfere unif} along with conditional i.i.d. property of $U$ with realization denoted by $u=\|\nbx_0+\nby_i\|$, and (c) follows from substituting the density function, $f_U(u|\nu_0)$, given by \eqref{eq:dis interfere unif}.  The final result can be obtained by using \cite[{eq(3.194.1)}]{zwillinger2014table}.
\subsection{Proof of Lemma~\ref{lem: k-closest}}
\label{App: proof of lemma Laplace k comm}
The Laplace transform of interference conditioned on $V_0$ and  $R=W_{k:N^{t}}$ is:
\begin{align*}
\ncalL_{\ncalI}^{(k)}(s|\nu_0,r)
&=\E\Big[\prod_{u_{\tt in} \in \ncalB^{\tt in}} \exp\Big(-s  h_i {u_{\tt in}}^{-\alpha}\Big) \prod_{u_{\tt out}\in \ncalB^{\tt out}}\exp\Big(s  h_i {u_{\tt out}}^{-\alpha}\Big)\Big]\\
&\stackrel{(a)}=\E\Big[\prod_{u_{\tt in}\in \ncalB^{\tt in}} \frac{1}{1+s   {u_{\tt in}}^{-\alpha}} \prod_{u_{\tt out}\in \ncalB^{\tt out}}\frac{1}{1+s   {u_{\tt out}}^{-\alpha}}\Big]\\
&\stackrel{(b)}= \sum_{\ell=0}^{n^{\rm a}_{m}}\frac{p^\ell (1-p)^{N^{\rm a}-\ell-1} \binom{N^{\rm a}-1}{\ell}} {\sum_{\ell=0}^{n^{\rm a}_{m}}p^\ell (1-p)^{N^{\rm a}-\ell-1} \binom{N^{\rm a}-1}{\ell}}\Bigg(\int_0^{r} \frac{1}{1+s   {u_{\tt in} }^{-\alpha}} f_{U_{\tt in}}(u_{\tt in}|\nu_0, r) \nrmd u_{\tt in} \Bigg)^{\ell} \\
&\times \Bigg(\int_r^{\nu_0+r_{\rm d}}\frac{1}{1+s   {u_{\tt out}}^{-\alpha}}  f_{U_{\rm out}}(u_{\tt out}|\nu_0, r) \nrmd u_{\tt out} \Bigg)^{N^{\rm a}-\ell-1}
\end{align*}
where $(a)$ follows from  $h_i \sim \exp(1)$, and $(b)$ follows from the fact that $u_{\tt in}\in \ncalB^{\tt in}$ and $u_{\tt out}\in \ncalB^{\tt out}$  are conditionally i.i.d. 
with PDF { $f_{U_{\tt in}}(.|\nu_0, r) $}, and  { $f_{U_{\rm out}}(.|\nu_0, r)$}, followed by the fact that number of nodes in ${\cal B}_{\tt in}$ is a binary random variable with probability $p=\frac{k-1}{N^{\rm t}-1}$ conditioned on total being less than $n^{\rm a}_{m}=\min(k-1,N^{\rm a}-1)$.  Substituting the PDFs of  $f_{U_{\tt in}}(.|\nu_0, r) $,  and  $f_{U_{\rm out}}(.|\nu_0, r)$ given by Lemma \ref{lem conditional_cdf_w}  completes the proof.
\subsection{Proof of Lemma~\ref{lem: joint success uniform}}
\label{App: proof of Lemma joint success uniform}

 According to the definition of joint success probability, we have 
  \begin{align*}
 &{\tt P}_{\rm joint}^{(u)}(\nu_0,n)=\P\big(\cap_{j=1}^n h_{\ell,j}>\beta r^{\alpha} {\cal I}_j\big)\stackrel{(a)}=\E\Big[\prod_{j=1}^n \exp(\beta r^{\alpha} {\cal I}_j)\Big]\\
 &=\E\Big[ \prod_{{\bf y}_i\in \Phi_{\rm a}\setminus {\bf y}_\ell} \prod_{j=1}^n  \exp\big(\beta r^{\alpha}   h_{i,j} \|{\bf x}_0+{\bf y}_i\|^{-\alpha}\big)\Big]
 \stackrel{(b)}=\E\Bigg[ \prod_{{\bf y}_i\in \Phi_{\rm a}\setminus {\bf y}_\ell} \bigg( \frac{1}{1+\beta r^{\alpha}   \|{\bf x}_0+{\bf y}_i\|^{-\alpha}}\bigg)^n\Bigg] \stackrel{(c)} =\E_R\bigg[\bigg(\\
 & \int_0^{r_{\rm d} - \nu_0} \left(\frac{ 1}{1+\beta r^{\alpha} u^{-\alpha}} \right)^n  \frac{2 u}{ r_{\rm d}^2} \nrmd u+
\int_{r_{\rm d} - \nu_0}^{r_{\rm d} + \nu_0}\left( \frac{1}{1+\beta r^{\alpha} u^{-\alpha}}\right)^n \frac{2 u}{\pi r_{\rm d}^2}\arccos \big(\frac{u^2+\nu_0^2-r_{\rm d}^2}{2 \nu_0 u}\big) \nrmd u \bigg)^{N^{\rm a}-1}\bigg]
 \end{align*}
where $(a)$ follows from $h_{\ell,j}\sim \exp(1)$, $(b)$ follows from  expectation over $h_{i,j}$-s along with the fact   $h_{i,j}$-s are i.i.d., and $(c)$ follows from converting  Cartesian to polar coordinates  by using the density function of distance given by \eqref{eq:dis interfere unif} along with conditional i.i.d. property of $u=\|\nbx_0+\nby_i\|$. The first integral of $(c)$ reduces to closed form expression by using  \cite[{eq (3.241.4)}]{zwillinger2014table} and some algebraic manipulation. The final result  is obtained by taking expectation over serving distance.

\bibliographystyle{IEEEtran}
\bibliography{BPP_Journal_Arxive.bbl}

\end{document}

%% file: notation.tex
\def\nbx{{\mathbf{x}}}
\def\nby{{\mathbf{y}}}

\def\nb0{{\mathbf{0}}}
\def\nb1{{\mathbf{1}}}

\def\ncalA{{\mathcal{A}}}
\def\ncalB{{\mathcal{B}}}

\def\ncalI{{\mathcal{I}}}

\def\ncalL{{\mathcal{L}}}

\def\nrmd{{\rm d}}

\newtheorem{lemma}{Lemma}
\newtheorem{thm}{Theorem}
\newtheorem{definition}{Definition}

\newtheorem{prop}{Proposition}
\newtheorem{cor}{Corollary}

\newtheorem{remark}{Remark}

\def\figref#1{Fig.\,\ref{#1}}%
\def\E{\mathbb{E}}

\def\P{\mathbb{P}}
\def\pc{\mathtt{P_c}}

\def\R{\mathbb{R}}

\def\T{\beta}							

\def\sir{\mathtt{SIR}}

